\documentclass[authoryear]{arxivnejsds}
\volume{0}
\issue{0}
\pubyear{2022}
\articletype{research-article}
\doi{0000}

\usepackage{times}
\usepackage{bm}
\usepackage{float}
\usepackage[]{algorithm2e}
\usepackage{amsthm}
\usepackage{amssymb}
\usepackage{natbib}
\usepackage[colorinlistoftodos]{todonotes}
\usepackage{mathtools}
\usepackage{framed}
\usepackage{csquotes}

\newcommand*{\inner}[2]{\langle #1, #2\rangle}
\newcommand*{\rme}{\mathrm{e}}
\newcommand*{\rmd}{\mathrm{d}}
\newcommand*{\PL}{\mathrm{PL}}
\newcommand*{\KL}{\mathrm{KL}}
\newcommand*{\bracks}[1]{\left\{#1\right\}}
\newcommand*{\sqbrack}[1]{\left[#1\right]}
\newcommand*{\paren}[1]{\left(#1\right)}

\newcommand*{\E}{\mathrm{E}}

\newcommand{\indicator}[1]{\mathbf{1} \{ #1 \} }

\newcommand*{\reals}{\mathbb{R}}

\DeclareMathOperator*{\argmax}{arg\,max}

\newtheorem{theorem}{Theorem}[section]
\newtheorem{corollary}[theorem]{Corollary}
\newtheorem{proposition}[theorem]{Proposition}

\theoremstyle{remark}

\theoremstyle{definition}

\theoremstyle{definition}

\theoremstyle{definition}

\theoremstyle{definition}
\newtheorem{lemma}[theorem]{Lemma}

\theoremstyle{definition}

\DeclareRobustCommand{\VOORVOEGSEL}[3]{#2} 

\begin{document}
\begin{frontmatter}
\pretitle{Research Article}
\title{The Anytime-Valid Logrank Test: Error Control Under Continuous Monitoring
  with Unlimited Horizon}

\begin{aug}
  \author[a]{
    \inits{J.}
    \fnms{Judith}
    \snm{ter Schure}%
    \thanksref{c1}
    \ead[label=e1]{j.a.terschure@amsterdamumc.nl}
  }
  \author[a]{
    \inits{M.F.}
    \fnms{Muriel F.}
    \snm{Pérez-Ortiz}%
    \thanksref{c1}
    \ead[label=e2]{muriel.perez@cwi.nl}
  }
  \author[a]{
    \inits{A.}
    \fnms{Alexander}
    \snm{Ly}
    \ead[label=e3]{a.ly@cwi.nl}
  }
  \author[a]{
    \inits{P.D.}
    \fnms{Peter D.}
    \snm{Grünwald}
    \ead[label=e4]{pdg@cwi.nl}
  }
  \thankstext[type=corresp,id=c1]{These authors share first authorship. Correspondence concerning this article may be addressed to xxxx Email address: YYY@ZZZZ.be.}
  \address[a]{Science Park 123, 1098XG Amsterdam \institution{Machine Learning
      Group, CWI}, \cny{The Netherlands}.\\ \printead{e1}}
  \address[b]{Science Park 123, 1098XG Amsterdam \institution{Machine Learning
      Group, CWI}, \cny{The Netherlands}.\\ \printead{e2}}
  \address[c]{Science Park 123, 1098XG Amsterdam \institution{Machine Learning
      Group, CWI}, \cny{The Netherlands}.\\ \printead{e3}}
  \address[a]{Science Park 123, 1098XG Amsterdam \institution{Machine Learning
      Group, CWI}, \cny{The Netherlands}.\\ \printead{e4}}
\end{aug}

\begin{abstract}
  We introduce the anytime-valid (AV) logrank test, a version of the logrank
  test that provides type-I error guarantees under optional stopping and
  optional continuation. The test is sequential without the need to specify a
  maximum sample size or stopping rule, and allows for cumulative meta-analysis
  with type-I error control. The method can be extended to define anytime-valid
  confidence intervals. The logrank test is an instance of the martingale tests
  based on $\E$-variables that have been recently developed. We demonstrate
  type-I error guarantees for the test in a semiparametric setting of
  proportional hazards and show how to extend it to ties, Cox' regression and
  confidence sequences. Using a Gaussian approximation on the logrank statistic,
  we show that the AV logrank test (which itself is always exact) has a similar
  rejection region to O'Brien-Fleming $\alpha$-spending but with the potential
  to achieve $100\%$ power by optional continuation. Although our approach to
  {\em study design\/} requires a larger sample size, the {\em expected\/}
  sample size is competitive by optional stopping.
\end{abstract}


\received{\smonth{10} \syear{2022}}

\end{frontmatter}

\section{Introduction}
\label{sec:introduction}
The logrank test is arguably the most important tool for the statistical
comparison of time-to-event data between two groups of participants. Our main focus
is when the two groups refer to the treatment and control groups in a randomized
controlled trial; the outcome of interest are event times, that is, the time elapsed until an outcome of
interest. The logrank test, in turn, uses a simplified version of the
proportional hazard ratio model of \citet{cox_regression_1972}. For a fixed
sample size and under this model, \citeauthor{cox_regression_1972} gave a simple
but profound insight: inference can be performed using the partial likelihood of
having observed the events in the particular order that they were observed. To
this end, the logrank test
\citep{mantel_evaluation_1966,peto_asymptotically_1972}, the score test
associated to the \citeauthor{cox_regression_1972}' partial likelihood, is
optimal for fixed sample size and a restricted alternative. Large-sample
properties of the logrank test are known in very general settings
\citep{tsiatis_large_1981,schoenfeld_asymptotic_1981,andersen_statistical_1993}.
Nevertheless, 
it is 
clear that the fixed-sample
assumption can be overly restrictive. Indeed, due to ethical and practical
constraints in human survival-time medical trials, interim analyses may be
performed to terminate the study earlier than planned if needed. Consequently,
it has been of fundamental importance to develop methods for the sequential
analysis of time-to-event data in general; for the logrank test, in particular.

In order to legitimate the use of sequential boundary decisions, uniform
asymptotic approximations over the study period have been developed for the
logrank statistic
\citep{tsiatis_group_1982,sellke_sequential_1983,slud_sequential_1984}. The
results in this line of work show the convergence of the sequentially computed
logrank statistic to a rescaled Brownian motion under very general censoring and
participant-arrival patterns. When interim analyses are only performed at discrete
times, the decision boundaries based on continuously monitoring the logrank
statistic are known to be overly conservative. This deficiency is addressed by
group-sequential and $\alpha$-spending methods, which, using knowledge of the
interim analysis times relative to a predefined maximum number of events,
allow for tighter decision boundaries
\citep{pocock_group_1977,obrien_multiple_1979,kim_confidence_1987}. These
sequential methods allow several interim looks at the data to stop for efficacy
(if the treatment shows to be beneficial) or futility (if the study is no longer
likely to reach statistical significance).

Despite the profound impact that these methods have had in statistical practice,
the requirement of a maximum sample size limits the utility of a promising but
nonsignificant study once the maximum sample size is reached. Because of their
design, extending such a trial makes it impossible to control their type-I
error. Moreover, the evidence gathered in new---possibly unplanned---trials
cannot be added in a typical retrospective meta-analysis, when the number of
trials or timing of the meta-analysis are dependent on the trial results. Such
dependencies introduce accumulation bias and invalidate the assumptions of
conventional statistical procedures in meta-analysis
\citep{terschure2019accumulation}. In order to address these deficiencies, we
look for flexible anytime-valid methods that provide type-I error control in two
situations: (1) optional stopping, which refers to halting the experiment
earlier or later than planned under arbitrary stopping rules, and (2)
meta-analysis and optional continuation, which refers to the aggregation of
evidence of possibly interdependent studies. Just as the existing methods, our
approach is connected to early work by H. Robbins and collaborators
\citep{darling_confidence_1967,lai_confidence_1976}. Most notably, existing
approaches come with fixed stopping rules, which are not desirable in the use
cases that are of our present interest. The details of the present approach are
very different, and to some extent, as we will see, more straightforward.

The main result of this work is the anytime-valid (AV) logrank test, an
anytime-valid test for the statistical comparison of time-to-event data from two
groups of participants. The AV logrank test uses the exact ratio of the sequentially
computed Cox partial likelihood as test statistic.  The advantage of having an
exact test manifests, for instance, in the case of unbalanced allocation, when
both control and treatment groups start with different numbers of participants.
In this case, $\alpha$-spending approaches do not provide strong type-I error
guarantees due to the approximations involved \citep{wu_group_2017}. The basic
version of the AV logrank test is, however, exact; unbalanced allocation
presents no difficulties. 

From a technical point of view, we show, under general patterns of incomplete
observation, that under the composite null hypothesis our test statistic is a
continuous-time martingale with expected value equal to one. Statistics with
this sequential property are referred to as test martingales; they form the
basis of anytime-valid tests \citep{ramdas_admissible_2020}. The AV logrank test
is a concrete instance of such a test martingale derived from the recent theory
of anytime-valid hypothesis testing based on $\E$-processes
\citep{henzi_valid_2021,grunwald_safe_2020,shafer_testing_2021,wang_false_2020}.
At each time an event takes place, it takes on the form of a likelihood ratio ratio between two multinomial distributions in a sampling-without-replacement setting. Relatedly, \citet{NEURIPS2022_12f3bd5d} consider discrete test martingales for multinomial experiments that can be interpreted in a sampling-with-replacement setting.  In contrast to $p$-values, an analysis based on $E$-processes can extend
existing trials as well as inform the decision to start new trials and
meta-analyses, while still controlling type-I error rate. Type-I error control
is retained even (i) if the $E$-process is monitored continuously and the trial
is stopped early whenever the evidence is convincing, (ii) if the evidence of a
promising trial is increased by extending the experiment and (iii) if a trial
result spurs a new trial with the intention to combine them in a meta-analysis.

The AV logrank test was developed with a specific application in mind and it
illustrates its usefulness. Some of the authors were involved in applying the AV
logrank test to the continuous meta-analysis of seven Coronavirus disease (COVID-19)  clinical trials ---the results are
available as a living systematic review 
 including code and summary data to reproduce the analysis \citep{terschure2022bcg}. This analysis was
performed concurrently with the trials in a so-called Anytime Live and Leading
Interim (ALL-IN) meta-analysis \citep{terschure2022all-in}. We
remark that even in the presence of dependencies between 
the existence and size of the
trials, the test based
on the multiplication of the values of $E$-processes retains type-I error
control as long as all trials test the same global null hypothesis, as was the
case in the above application. This is generally useful if we want to combine
the results of several trials in a bottom-up retrospective meta-analysis, where
no top-down stopping rule can be enforced. It is even possible to obtain an interim meta-analysis result by combining
{\em interim\/} results of {\em ongoing\/} trials by multiplication, stepping beyond the realm
of existing sequential approaches.

\subsection{Contributions and outline}
\label{sec:contr-outl}
We begin with Section~\ref{sec:model}, where we review the special instance Cox'
proportional hazards model for the two-group setting. There, we set the
assumptions and notation used in the rest of the article. The definitions
presented there are standard.
In Section~\ref{sec:safe-logrank-test-1}, we define and prove that the AV
logrank test is indeed anytime valid. We first do this for (a) the case with
only a group indicator (no other covariates) and without simultaneous events
(ties). There, we also discuss its optimality properties and extend it to (b)
the case with ties and to (c) the case when one wants to learn the actual effect
size of the data and/or use prior knowledge about the effect size  via a Bayesian prior.
The
resulting version of the test keeps providing nonasymptotic type-I error control
even if the priors are wildly misspecified, that is, if they predict very
different data from the data we actually observe. These results hinge on showing
that the likelihood underlying Cox' proportional hazards model can be used to
define $E$-variables and test martingales.
In Section~\ref{sec:appr-safel-test}, we show a Gaussian approximation to the
AV logrank statistic that is useful in the common situation when only summary
statistics are available.
We then provide extensive computer simulations to compare the AV logrank test to
the classic logrank test and $\alpha$-spending approaches.
In Section~\ref{sec:rejection-regions}, we show that the exact AV
logrank test has a similar rejection region to O'Brien-Fleming $\alpha$-spending
for those designs and hazard ratios where it is well-approximated by a Gaussian
AV logrank test. While always needing a small amount of extra data in the
design phase (the price for indefinite optional continuation), the expected
sample sizes needed for true rejections remain very competitive. During the design
phase of a study, we might want to design for a maximum sample size in order to
achieve a certain power, but need a smaller sample size on average during the
study since we can safely engage in optional stopping.
In Section~\ref{sec:optional-continuation}, we show that AV-logrank-type tests
can be combined through multiplication to perform meta-analysis, and in
Section~\ref{sec:confidence-sequences}, we show how the test can be used to
derive confidence sequences for the hazard ratio.
In Section~\ref{sec:power-sample-size}, we compare the sample sizes that are
needed during the design phase in order to achieve a targeted power.
Lastly, in Section~\ref{sec:disc-concl-future} we make concluding remarks and
discuss future research directions.

We remark that once the definitions are in place, the technical results are
mostly straightforward consequences from earlier work; in particular, of the
work of \citet{cox_partial_1975,slud_partial_1992} and
\citet{andersen_statistical_1993}. The novelty of the present work is thus
mainly in {\em defining\/} the AV logrank test and showing by computer
simulation that, while being substantially more flexible, it is competitive with
existing approaches---the classic logrank test with fixed design and in
combination with $\alpha$-spending.

Next to the main body of this article, we provide two appendices. We delegate to
Appendix~\ref{sec:math} proofs and remarks that, while important, are not
needed to follow the main development. Most importantly, the particular
$\E$-variable we design is \textit{growth-rate optimal in the worst case}, GROW
(see Section~\ref{sec:optimality}). \citet{grunwald_safe_2020} provide several
motivations for this criterion; we provide an additional one using an argument
of \citet{breiman_optimal_1961}, which does not seem to be widely known. This
argument shows a connection between growth-rate optimality and tests with
minimal expected stopping time. In Appendix~\ref{app:extensions}, we provide an
extension to the case when covariates other than group membership are present.
This extension, based on the full Cox model, requires solving a challenging
optimization problem and its implementation is therefore deferred to future
work.

\section{Proportional hazards model and Cox' partial likelihood}
\label{sec:model}
We begin by describing the hypothesis that is being tested, the data that are
available, and \citeauthor{cox_regression_1972}' proportional hazards model. We
are interested in comparing the survival rates between two groups of participants,
Group $A$ and Group $B$. In a randomized controlled trial, Group $A$ would signify
the control group; Group $B$, the treatment group. We assume that the available
data about $m$ participants are of the form
$\{ (X^i, g^i, \delta^i): i = 1,\dots, m \}$, where $X^i = \min\{T^i, C^i\}$ is
the minimum between the event time $T^i$ and the (possibly infinite) censoring
time $C^i$; $g^i$ is a zero-one covariate depending on group membership
($g^i = 0$ signifies that $i \in A$; $g^i = 1$, that $i \in B$); and
$\delta^i = \indicator{X^i = T^i}$ is the indicator of whether the event was
witnessed before censoring or not. Let $m^A$ be the number of members of Group $A$
and $m^B$ the number of members of Group $B$---then $m^A +m^B = m$. Define
$\mathbf{g} = (g^1, \dots, g^n)$, the vector of group memberships. We assume
that $T^1, \dots, T^n, C^1,\dots, C^n$ are independent and have continuous
distribution functions. The continuity assumption precludes tied observations;
we relax this assumption later on, in Section~\ref{sec:ties-1}. For
$i = 1, \dots, m$, the survival rates are quantified by the hazard functions
$\lambda^i = (\lambda_t^i)_{t \geq 0}$ for $T_i$, given by %
\begin{equation}
\label{eq:definition_hazard_function}
  \lambda^i_t = -\frac{\rmd}{\rmd t} \ln \mathbf{P}\{T^i \geq t\}.
\end{equation}
As is customary, the hazard function $\lambda^i$ at \( t \) can be interpreted via the
conditional probability of witnessing an event in a short time span provided
that the event has not been witnessed up to \( t \), that is, %
\begin{equation}
\label{eq:cox_conditional_jump_probability}
  \mathbf{P}\{t \leq T^i < t + \Delta t  \ | \ t \leq T^i \} =  \lambda^i_t \Delta t+ o(\Delta t)
  \text{ \ as \ }
  \Delta t \to 0.
\end{equation}
Given our interest in comparing the survival rates between the two groups,
suppose that all participants $i$ of Group $A$ have a common hazard function
$\lambda^i_t = \lambda^{A}_t$; members $i$ of Group $B$,
$\lambda^i_t = \lambda^{B}_t$. Using the data, we wish to test proportional
hazards hypotheses. Concretely, we test the hypotheses ${\cal H}_0$ that the
hazard function of the members of both groups satisfy
$\lambda^A_t = \theta_0\lambda_t^B$, against an alternative hypothesis
${\cal H}_1$ that \( \lambda^{B}_{t} = \theta \lambda^{A}_{t} \) for a
\( \theta \neq \theta_0 \). As a first application of the methods that we
develop, we consider the statistical hypothesis testing problem between the null
hypothesis that the hazard functions of the two groups are the same against the
left-sided alternative, that is,
\begin{equation}
\label{eq:general_testing_problem}
\begin{multlined}
  {\cal H}_0: \lambda^{B}_t = \theta_0\lambda^{A}_t \text{ \ \ vs. \ \ } {\cal
    H}_1: \lambda^{B}_t = \theta\lambda^{A}_t\\ \text{\ \ for some \ } \theta
  \leq \theta_1 < \theta_0 \text{ and all } t,
\end{multlined}
\end{equation}
where \( \theta \) is known as the hazard ratio and is the main quantity of
statistical interest, and \( \theta_1 \) would be, in a clinical trial, a
minimal clinically relevant effect size. The alternative is what we hope for in
case of negative events, such as death, with treatments that are set out to
lower (relative to the control condition) the hazard rate. Notice that the
hypotheses in \eqref{eq:general_testing_problem} are, in fact, nonparametric.
Similarly, if the event is positive, e.g., recovery from an infection, we would
typically set a right-sided alternative, which can be also be treated with the
present methods.

Right-sided, two-sided and the full alternative hypothesis
$\mathcal{H}_1':\theta\neq 1$ are also amenable to the methods that will follow.
We remark, however, that all the methods retain their type-I error guarantees
irrespective of the specific alternative that we use.
%
%
We now turn to defining Cox' partial
likelihood $\PL_t$, which is at the center of our approach. To that end, we
need a battery of standard definitions---we lay them out to establish the
notation. Let $y^i_t = \indicator{X^i\geq t}$ be the at-risk process, that is,
the indicator of whether participant $i$ is still at risk at time $t$, and let
$\bar{y}^A_t = \sum_{i \in A} y^i_t$ and $\bar{y}^B_{t} = \sum_{i \in B}y^i_t$ be
the number of participants at risk in each of the groups at time $t$. Define
$\mathbf{y}_t = (y^1_t, \dots, y^m_t)$, the vector of at-risk processes, and
${\cal R}_t = \{j: y^j_t = 1\}$, the set of participants at risk at time $t$. Let
$T^{(1)} < T^{(2)} < \dots < T^{(\bar{N}_{\infty})}$ be the set of ordered events
times that were witnessed (not censored). Note that, if all participants witness the
event and censoring is absent, $\bar{N}_{\infty} = m$. For each
$k = 1, \dots, \bar{N}_{\infty}$, let $I_{(k)}$ be the index of the individual
that witnessed the event at time $T^{(k)}$. This means, for example, that if
participant with label three was the fifth to witness the event, then $I_{(5)} = 3$.
Abbreviate by $y^i_{(k)}, \bar{y}^A_{(k)}, \bar{y}^B_{(k)}, {\cal R}_{(k)}$ the
corresponding quantities at event time $T^{(k)}$, and define
$g^{(k)}:= g^{I_{(k)}}$. \citeauthor{cox_regression_1972}' partial likelihood
$\PL_{\theta, t}$ can be sequentially computed by %
\begin{equation}
\label{eq:def_cox_partial_likelihood}
  \PL_{\theta, t}
  =
  \prod_{k : T^{(k)} \leq t}
  \frac{
    \theta^{g^{(k)}}
  }{
    \sum_{l\in \mathcal{R}_{T^{(k)}}}
    \theta^{g^{l}}
  }
  =
  \prod_{k : T^{(k)} \leq t}
  \frac{\theta^{g^{(k)}}}{\bar{y}^A_{(k)} + \theta \bar{y}^B_{(k)}}.
\end{equation}
Cox' likelihood evaluated at the event times $T^{(1)}, T^{(2)}, \dots$
coincides to that of a sequence of multinomial trials where, at event time
$T^{(k)}$, each of the participants $i \in \mathcal{R}_{(k)}$ witnesses the event with
probability %
\begin{align}
  &p_{\theta, (k)}( \ i \ )
    := \mathbf{P}\{
    I_{(k)} = i
    \mid
    \mathbf{y}_{(l)}, \mathbf{g};  \ l = 1, \dots k
    \},   \nonumber \\
  &\label{eq:likelihood_per_time}
  p_{\theta, (k)}( \ i \ )
    =
    \frac{\theta^{g^{i}} }{\bar{y}^A_{(k)} + \theta
    \bar{y}^B_{(k)}}.
\end{align}
\citeauthor{cox_regression_1972} showed that, indeed, conditionally on all the
information accrued strictly before $T^{(k)}$, the probability that participant $i$
observes an event at time $T^{(k)}$ is exactly $p_{\theta,(k)}( \ i \ )$ as long
as the hazard ratio is $\theta$. With these likelihood computations at hand, we
are in place to show the main contribution of this article, the AV logrank
test, which uses the partial likelihood ratio as the test statistic.

\section{The AV logrank test}
\label{sec:safe-logrank-test-1}
In this section the AV logrank test for~\eqref{eq:general_testing_problem} is
introduced; its type-I error guarantees and optimality properties are
investigated. We give a solution to the first of the purposes laid down in the
introduction: we show that the AV logrank test is anytime valid---its type-I
error guarantees are not affected by optional stopping. The fact that it is also
type-I-error-safe under optional continuation, our second purpose, is proven in
Section~\ref{sec:optional-continuation}. %
Without further ado, we define the AV
logrank statistic $S_{\theta_0, t}^{\theta_1}$, typically, \( \theta_0 = 1 \),
for~\eqref{eq:general_testing_problem} as the partial likelihood ratio %
\begin{equation}\label{eq:def_safe_logrank_statistic}
  S_{\theta_0, t}^{\theta_1}
  =
  \frac{\PL_{\theta_1,t}}{  \PL_{\theta_0,t}}
  =
  \prod_{k: T^{(k)} \leq t} \frac{p_{\theta_1, (k)}(I_{(k)})}{p_{\theta_0, (k)}(I_{(k)})}.
\end{equation}
Here, $p_{\theta,(k)}$ is as defined in \eqref{eq:likelihood_per_time}; the
product that defines our statistic $S_{\theta_0,t}^{\theta_1}$ runs over the
events that have been witnessed up to and including time $t$, and the empty
product is taken to be equal to one. As is conventional with likelihood ratios,
high values of $S^{\theta_1}_{\theta_0,t}$ are indicative that the alternative
hypothesis is better than the null hypothesis at the describing the data. Given
a tolerable type-I error bound $\alpha$ and an arbitrary random time $\tau$, the
AV logrank test is the test that rejects the null hypothesis if
$S_{\theta_0\tau}^{\theta_1}$ is above the threshold $1/\alpha$, that is, %
\begin{equation}
\label{eq:def_safe_logrank_test}
  \xi_{\theta_0,\tau}^{\theta_1} = \indicator{S_{\theta_0,\tau}^{\theta_1} \geq 1 / \alpha}
  :=
  \begin{cases}
    1 & \text{ if } S_{\theta_0, \tau}^{\theta_1} \geq 1 / \alpha \\
    0 & \text{ otherwise}.
  \end{cases}
\end{equation}
As we will see, by its sequential properties, $S_{\theta_0,t}^{\theta_1}$ takes
large values with small probability under the null hypothesis uniformly over
time, which translates into type-I error control for the test
$\xi_{\theta_0, \tau}^{\theta_1}$. This observation is behind the any-time
validity of the AV logrank test, and of anytime-valid tests in general (more
details and general constructions to the effect of anytime-valid sequential
testing can be found in the work of \citet{ramdas_admissible_2020}). We shown in
the following proposition that the test $\xi_{\theta_0,\tau}^{\theta_1}$ has the
desired type-I error control.
\begin{proposition}
\label{prop:anytime-valid}
  Let $\mathbf{P}_0$ be any distribution under which the hazard ratio is equal
  to $\theta_0$, and let $\tau$ be any random time. The test
  $\xi_{\theta_0,\tau}^{\theta_1} = \indicator{S_{\theta_0, \tau}^{\theta_1}
    \geq 1 / \alpha}$, where $S_{\theta_0, t}^{\theta_1}$ is as in
  (\ref{eq:def_safe_logrank_statistic}), has level $\alpha$, that is,
  \begin{equation*}
    \mathbf{P}_0\{\xi_{\theta_0, \tau}^{\theta_1} = 1 \} \leq \alpha.
  \end{equation*}
\end{proposition}
This result can be readily obtained using the sequential-multinomial
interpretation of Cox' likelihood ratio. As we will see, in
Section~\ref{sec:optimality}, this result can be interpreted in terms of
$E$-variables and $E$-processes \citep{grunwald_safe_2020}. Define the process
$(S_{\theta_0, (k)}^{\theta_1})_{k = 1, 2,\dots}$ as the value of the AV
logrank statistic at the event times $T^{(k)}$, that is,
$S_{\theta_0, (k)}^{\theta_1} := S_{\theta_0, T^{(k)}}^{\theta_1}$. In this time discretization, the
AV logrank statistic is the product of random variables %
\begin{equation}
\label{eq:one-outcome-evariable}
  R_{\theta_0, (k)}^{\theta_1} = p_{\theta_1,(k)}(I_{(k)}) /
  p_{\theta_0,(k)}(I_{(k)}),
\end{equation}
the one-outcome partial likelihood ratio for the
$k$th event, where $p_{\theta_0,(k)}$ is as in \eqref{eq:likelihood_per_time}
and $k = 1, 2, \dots$.

\begin{proof}[Proof of Proposition~\ref{prop:anytime-valid}]
  Under any distribution under which the hazard ratio is $\theta_0$, the fact
  that the likelihood of observing $I_{(k)}$ conditionally on
  $\{\mathbf{y}_{(l)}: l = 1, \dots, k\}$ equals $p_{\theta_0,(k)}(I_{(K)})$
  implies that
  \begin{equation}
  %
  \label{eq:one-outcome-evariable2}
    \mathbf{E}[
    R_{\theta_0, (k)}^{\theta_1}
    \mid
    \mathbf{y}_{(1)}, \dots, \mathbf{y}_{(k)}
    ]
    =
    \sum_{j\in \mathcal{R}_{(k)}}
    p_{\theta_0,(k)}( j )
    \frac{
      p_{\theta_1,(k)}( j )
    }{
      p_{\theta_0,(k)}(j)
    } = 1.
  \end{equation}
  This immediately shows that
  $S_{\theta_0,(k)}^{\theta_1}= \prod_{i\leq k} R_{\theta_0,(k)}^{\theta_1}$ is
  a test martingale, a nonnegative martingale with expected value equal to one,
  with respect to the filtration
  $\mathcal{F}_{-} = (\mathcal{F}_{(k)^-})_{k = 1,2, \dots}$ of sigma-algebras
  $\mathcal{F}_{(k)^-} = \sigma(\mathbf{y}_{(k)}: k = 1,\dots, k)$. Next, the
  type-I error control for the the test $\xi_{\theta_0}^{\theta_1}$ follows from
  Ville's inequality, which asserts that, under the null hypothesis, the test
  martingale $S_{\theta_0, (k)}^{\theta_1}$ takes large values with small
  probability. Ville's inequality \citep{ville_etude_1939} (also known as Doob's
  maximal inequality) implies that
  \begin{equation*}
    \mathbf{P}\bracks{
      \sup_{k = 1, 2, \dots} S_{\theta_0, (k)}^{\theta_1} \geq 1 / \alpha
    }
    \leq
    \mathbf{E}[S_{\theta_0, (1)}^{\theta_1}]
    \alpha
    =
    \alpha.
  \end{equation*}
  The previous display is a bound on ever making a type-I error when using the
  AV logrank test $\xi_{\theta_0,\tau}^{\theta_1}$.
\end{proof}

Under general patterns of incomplete observation---like independent censoring or
independent left truncation---, the AV logrank test provides the same type-I
error guarantees. To proof this, we give an alternative proof of
Proposition~\ref{prop:anytime-valid} in Appendix~\ref{sec:math} using the
counting-process formalism \citep{andersen_statistical_1993}. There, we show
that if the compensators of the underlying counting processes have a certain
general product structure---which is the case under complete observation---, the
AV logrank test is anytime-valid. We then refer to
\citet{andersen_statistical_1993}, who show that this structure is preserved
under said patterns of incomplete observation.

The AV-logrank test is optimal---in a sense to be defined in the next
section---among a large family of statistics. A second look at the proof of
Proposition~\ref{prop:anytime-valid} suggests a generalization of the AV logrank
statistic given in (\ref{eq:def_safe_logrank_statistic}). Let, for each $k$,
$q_{(k)}$ be a probability distribution on participants in the risk set
$\mathcal{R}_{(k)}$ which is only allowed to depend on
$\mathbf{y}_{(1)}, \dots,\mathbf{y}_{(k)}$. Analogously to
\eqref{eq:one-outcome-evariable}, we define the one-outcome ratio
$R_{\theta_0, (k)}^{q} := q_{(k)}(I_{(k)}) / p_{\theta_0,(k)}(I_{(k)})$---we now
use $q_{(k)}$ instead of $p_{\theta_1}$---, and
\begin{equation}
\label{eq:definition_generalized_statistic}
  S_{\theta_0, t}^q
  := \prod_{k: T^{(k)} \leq t} R_{\theta_0, (k)}^{q}
  = \prod_{k: T^{(k)} \leq t} \frac{q_{(k)}(I_{(k)})}{p_{\theta_0,(k)}(I_{(k)})}.
\end{equation}
A modification of the previous argument shows, for any random time $\tau$, a
type-I error guarantee for the test $\xi_{\theta_0, \tau}^q$ based on the value
of $S_{\theta_0,\tau}^q$, that is,
$\xi_{\theta_0,\tau}^q := \indicator{S_{\theta_0,\tau}^q \geq 1 / \alpha}$ (see
Proposition~\ref{prop:anytime-valid}). Any such test is also anytime valid as
long as each $q_{(k)}$ depends on the data only through
$\mathbf{y}_{(1)}, \dots, \mathbf{y}_{(k)}$. In Section~\ref{sec:impr-altern},
we use this generalization to provide tests when no value of $\theta_1$ is
available. This generalization raises a natural question about the optimality of
the AV logrank test based on (\ref{eq:def_safe_logrank_statistic}) among test
statistics of the form (\ref{eq:definition_generalized_statistic}). This is the
subject of the next section.

\subsection{E-variables and optimality}
\label{sec:optimality}
The random variables $\{R_{\theta_0, (k)}^{\theta_1}\}_{k = 1, 2\dots}$
from~\eqref{eq:one-outcome-evariable} and
$\{R_{\theta_0, (k)}^{q}\}_{k = 1, 2\dots}$ from
\eqref{eq:definition_generalized_statistic} are examples of
(conditional) $E$-variables---nonnegative random variables whose (conditional) 
expected value is below 1 uniformly over the null hypothesis. $E$-variables and
$E$-processes are the ``correct'' generalization of likelihood ratios to the
case that either or both $\mathcal{H}_{0}$ and $\mathcal{H}_{1}$ are composite
and can be interpreted in terms of gambling
\citep{grunwald_safe_2020,shafer_testing_2021,ramdas_admissible_2020}. Under
this gambling interpretation, a test martingale, a product of conditional
$E$-variables, is the total profit made in a sequential gambling game where no
earnings are expected under the null hypothesis. The analogy is thus between
profit and evidence: no evidence can be gained against the null hypothesis if it
is true. Just as $p$-values, the definition of $E$-variables and test
martingales does not need any mention of an alternative hypothesis. However, if
a composite set of alternative distributions is available, a gambler who is
skeptical of the null distribution might want to maximize the speed of evidence
accumulation (or of capital growth) under the alternative hypothesis. The
worst-case growth rate is defined (conservatively) as the smallest expectation
of the logarithm of the $E$-variable under the alternative. Consequently, any
$E$-variable achieving it is called GROW, for Growth-Rate Optimal in the Worst
case (see the work of \citet{grunwald_safe_2020} and \citet{shafer_testing_2021}
for additional reasons to use this optimality criterion).

We instantiate this reasoning to our present problem. For
the left-sided alternative \eqref{eq:general_testing_problem}, the choice
$R^{\theta_1}_{\theta_0, (k)}$ is conditionally GROW because it maximizes the
worst-case conditional growth rate
\begin{equation*}
  R^{q}_{\theta_0, (k)}
  \mapsto
  \min_{\theta \leq \theta_1}
  \mathbf{E}_{\theta}[ \ln R^{q}_{\theta_0, (k)} | \mathbf{y}_{(1)}, \dots, \mathbf{y}_{(k)}],
\end{equation*}
over all valid choices of $q_{(k)}$ (which can only depend on the data through
$\mathbf{y}_{(1)}, \dots, \mathbf{y}_{(k)}$), that is,
\begin{equation*}
  \begin{multlined}[b]
    \min_{\theta \leq \theta_1} \mathbf{E}_{\theta}[ \ln R^{\theta_1}_{\theta_0,
      (k)} | \mathbf{y}_{(1)}, \dots, \mathbf{y}_{(k)}]
    \\
    = \max_{q} \min_{\theta \leq \theta_1} \mathbf{E}_{\theta}[ \ln
    R^{q}_{\theta_0, (k)} | \mathbf{y}_{(1)}, \dots, \mathbf{y}_{(k)}].
  \end{multlined}
\end{equation*}
In Appendix~\ref{app:math-optimality}, we show that in the limit that the risk
sets are much larger than the number of events that are witnessed, this
worst-case growth criterion yields a test that minimizes the worst-case expected
stopping time---under the alternative hypothesis---among the tests that stop as
soon as $S_{\theta_0, t}^{q}\geq 1 / \alpha$. Thus, among all possible AV
logrank tests of the form~(\ref{eq:definition_generalized_statistic}), there are
strong reasons to choose $\xi^{\theta_1}_{\theta_0,\tau}$.

In a similar fashion, a test can be constructed for two sided alternatives.
Indeed, consider a testing problem of the form
\begin{equation}
  \label{eq:two-sided_testing_problem}
  \begin{split}
    {\cal H}_0: & \lambda^{B} = \lambda^{A}
    \text{ \ \ vs. \ \ } \\
    {\cal H}_1: & \lambda^{B} = \theta\lambda^{A} \text{\ \ for some \ } \theta
    \leq \theta_1
    \text{ \ \ or \ }
    \theta \geq 1 / \theta_1,
  \end{split}
\end{equation}
where $\theta_1 < 1$. For this problem, we can create a weighted, conditionally
GROW, $\E$-variable by using
$R^{2-\mathrm{sided}} = \frac{1}{2}R^{\theta_1}_{\theta_0, (k)} +
\frac{1}{2}R^{1 / \theta_1}_{\theta_0, (k)}$.

\subsection{Learning the hazard ratio from data}
\label{sec:impr-altern}
So far, the alternative hypotheses that we have studied are of the form
${\cal H}_1: \theta \leq \theta_1$ for some value of $\theta_1 < 1$. In some
cases, such a value of $\theta_1$ is available from the context of the analysis.
For instance, $\theta_1$ can correspond to a minimal clinically relevant effect
that is satisfactory in a medical trial. However, sometimes it is not clear which
value $\theta_1$ to chose. Still, statistics of the form
(\ref{eq:definition_generalized_statistic}) are useful to test a null hypothesis
${\cal H}_0$ as in~(\ref{eq:general_testing_problem}). Indeed, for each $k$, we
can use conditional probability mass functions $q_{(k)}$ that depend on data
observed on $t< T^{(k)}$ and enable us to implicitly learn the hazard ratio
$\theta$. We describe two such alternatives: a prequential plug-in likelihood
and Bayes predictive distribution.

\subsubsection{Prequential plugin test approach}
Using only the data observed in $t< T^{(k)}$, let $\hat{\theta}_{(k)}$ be the
smoothed maximum likelihood estimator
\begin{equation*}
  \hat{\theta}_{(k)}
  =
  \argmax_{\theta \geq 0}
  \paren{p_{\theta, 0}\times \prod_{k : T^{(k)} < t} p_{\theta, (k)}(I_{(k)})},
\end{equation*}
where $p_{\theta,0}$ is a smoothing based on the likelihood of having observed
two ``virtual'' data points prior to the observed data, that is,
$p_{\theta,0} = 1/(\bar{y}^A_0 + 1 + \theta(\bar{y}^B_0 + 1)) \times \theta /
(\bar{y}^A_0 + \theta(\bar{y}^B_0 + 1))$. The statistic
$S^{\mathrm{preq}}_{\theta_0, t}$ is (\ref{eq:definition_generalized_statistic})
with $q_{(k)} = p_{\hat{\theta}_{(k)}, (k)}$, and it can also be used to define
an anytime-valid test. With this choice, the process $q_{(1)}, q_{(2)}\dots, $
is a typical instance of a {\em prequential plug-in\/} likelihood
\citep{dawid_present_1984}, that is often based on suitable smoothed
likelihood-based estimators \citep{grunwald_minimum_2019}. The rationale behind
this method is the following. Suppose the data are actually sampled from a
distribution according to which the hazard ratio is $\theta$. For sufficiently
large initial risk sets, that is, if $\bar{y}^A_0$ and $\bar{y}^B_0$ are not too
small, by the law of large numbers, the smoothed maximum likelihood estimate
$\hat\theta_{(k)}$ will with high probability be close to $\theta$. Therefore,
$p_{\hat\theta, (k)}$ will behave more and more like the real $p_{\theta, (k)}$
from which data are sampled. Thus, the process $S^{\mathrm{preq}}_{\theta_0}$,
will behave more and more similarly to the ``correct'' partial likelihood ratio
(\ref{eq:def_safe_logrank_statistic}).

\subsubsection{Bayesian approach}
Instead of $q_{(k)}$ based on a plug-in estimate of $\theta$, it is also
possible to use a Bayes predictive distribution based on a prior $\mathbf{W}$ on
$\theta$. If
$\mathbf{W}_{(k)} = \mathbf{W} \ | \ \mathbf{y}_{(1)}, \dots, \mathbf{y}_{(k)}$
is the Bayes posterior on $\theta$ based on a prior $\mathbf{W}$ and the data up
to time $t < T^{(k)}$, then
\begin{equation*}
  q_{(k)} = p_{\mathbf{W},(k)} := \int p_{\theta, (k)} \rmd\mathbf{W}_{(k)}(\theta),
\end{equation*}
where $\mathbf{W}_{(1)} = \mathbf{W}$. Hence, $p_{\mathbf{W}, (k)}$ is the
Bayesian predictive distribution. The resulting statistic $S^{\mathbf{W}}_{t} $
is the result of multiplying the conditional probability mass functions
$p_{\mathbf{W},(k)}$, and we obtain that
\begin{equation}
\label{eq:bayesian}
  S^{\mathbf{W}}_{\theta_0, t} = \prod_{k:T^{(k)} \leq t}^n \frac{p_{\mathbf{W},(k)}(I_{(k)}) }{ p_{\theta_0, (k)}(I_{(k)})}
\end{equation}
is a Bayes factor between the Bayes marginal distribution based on $\mathbf{W}$
and $\theta_0$. This technique has been employed in sequential analysis; it is
known as the method of mixtures
\citep{darling_confidence_1967,robbins_boundary_1970}. We do not know of a prior
for which \eqref{eq:bayesian} or the constituent products have an analytic
expression, but it can certainly be implemented using, for example, Gibbs
sampling.

As shown in Section~\ref{sec:safe-logrank-test-1}, the use of any $S^{q}_{\theta_0,t}$
instead of $S^{\theta_1}_{\theta,t}$ does not compromise on safety: a test based
on monitoring $S^{q}_{\theta_0}$ is anytime-valid, whether $q$ makes reference
to plug-in estimators or Bayes predictive distributions, no matter what prior
$\mathbf{W}$ was chosen. The type-I error guarantee always holds, also when the
prior is ``misspecified'', putting most of its mass in a region of the parameter
space far from the actual $\theta$ from which the data were sampled. Thus, our
set-up is intimately related to the concept of {\em luckiness\/} in the machine
learning theory literature \citep{grunwald_tight_2019} rather than to ``pure''
Bayesian statistics. Indeed, given a target value $\theta_1$---a minimal
clinically relevant effect size---the \emph{worst-case} logarithmic growth rate of
$S^{q}_{\theta_0,t}$ will in general be smaller than that of the GROW
$S^{\theta_1}_{\theta_0,t}$. Nevertheless, $S^{q}_{\theta_0,t}$ can come
close to the optimal for a whole range of potentially data-generating $\theta$
and may thus sometimes be preferable over choosing
$S^{\theta_1}_{\theta_{0},t}$. More precisely, the use of a prior allows us to
exploit favorable situations in which $\theta$ is even smaller (more extreme)
than $\theta_{1}$. In such situations, the GROW $S_{\theta_{0},t}^{\theta_1}$ is
effectively misspecified. By using $S^q_{\theta_0,t}$ that learn from the data,
we may actually obtain a test martingale that grows faster than the GROW
$S^{\theta_1}_{\theta_0,t}$, which is fully committed to detecting the
worst-case $\theta_1$.

\begin{figure*}[t!]
  \centering
  \includegraphics[scale = 1]{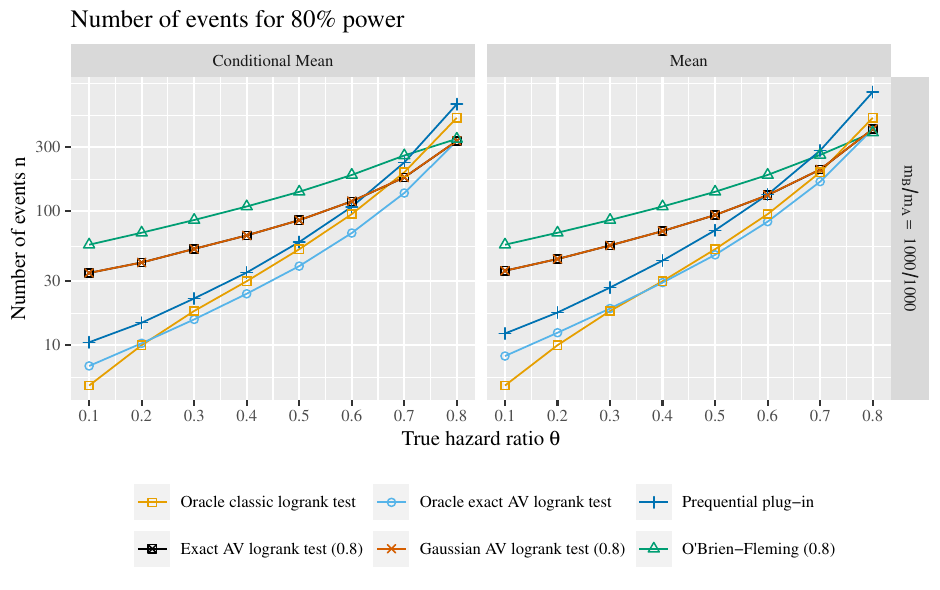}
  \caption{We show the number of events at which one can stop retaining 80\%
    power at $\alpha = 0.05$ using the process $S_{\theta_0,t}^{\theta_1}$ with
    $\theta_0 = 1$ and $\theta_1= 0.80$ when the true hazard ratio $\theta$
    generating the data are different from $\theta_1$. ``Oracle'' means that the
    method is specified with knowledge of the true $\theta$, which in reality is
    unknown. Note that the y-axis is
    logarithmic.\label{fig:simulation_sampleSize_misspecified} }
\end{figure*}

In \autoref{fig:simulation_sampleSize_misspecified}, we illustrate such a
situation where we start with 1000 participants in both groups. We generated
data using different hazard ratios, and used a `misspecified'
$S^{\theta_1}_{\theta_0,t}$ that always used $\theta_1 = 0.8$. Note that while
this is still the GROW (minimax optimal) test martingale for
$\mathcal{H}_1: \theta \leq \theta_1 \leq 0.8$. If we knew the true $\theta$, we
could use the test martingale $S_{\theta_0,t}^{\theta}$---it grows faster. We
will call the test based on this latter martingale the {\em oracle\/} exact AV
logrank test because it is based on inaccessible (oracle) knowledge. We
estimated the number of events needed to reject the null with 80\% power for
$S_{\theta_0,t}^{0.8}$, the oracle $S_{\theta_0,t}^{\theta}$, and the
prequential plug-in $S_{\theta_0,t}^{\mathrm{preq.}}$. In all cases, we used the
aggressive stopping rule that stops as soon as the statistic in question crosses
the threshold $1/\alpha = 20$. We see that, as the true $\theta$ gets smaller
than $0.8$, we need fewer events using the GROW test $S^{0.8}_{\theta_0,t}$ (the
data are favorable to us), but using the oracle exact AV logrank test we get a
considerable additional reduction. The prequential plug-in
$S_{\theta_0}^{\mathrm{preq.}}$ `tracks' the oracle $S_{\theta_0,t}^{\theta}$ by
learning the true $\theta$ from the data: for $\theta$ near $0.8$, it behaves
worse (more data are needed) than $S_{\theta_0,t}^{0.8}$ (which knows the right
$\theta$ from the start), but for $\theta < 0.6$ it starts to behave better. For
comparison we also added the methods discussed in
Section~\ref{sec:rejection-regions}.
Notably, the O'Brien-Fleming procedure,
even though unsuitable for optional continuation, needs even more events than
the misspecified AV logrank test $S_{\theta_0,t}^{0.8}$ as soon as $\theta$
goes below $0.8$. The simulations were performed using exactly the same
algorithms as for \autoref{fig:simulation_expected_vs_worst} so the $y$-axis at
$\theta = 0.8$ coincides with that of
\autoref{fig:simulation_expected_vs_worst}, but now with absolute rather than
relative numbers; details are described in Appendix~\ref{app:moreSampleSize}.

\subsection{Tied observations}
\label{sec:ties-1}
Here, we propose a sequential test for applications where events are not
monitored continuously, but only at certain observation times. In this case,
more than one event may be witnessed in the time interval between two
observation moments. Since the order in which these observations are made would
be unknown, our previous approaches fail to offer a satisfactory sequential
test. Assume that we make observations at times $t_0 < t_1 < t_2 < \dots$ that
are fixed before the start of the study. Even though we assume the absence of
censoring in this section, this approach can be adapted to its presence under an
additional common assumption: that the events reported between two observation
times $t_{k-1}$ and $t_k$ precede any censorings, so that censored patients
contribute fully to the risk sets under consideration. We assume that the
available data are of the form $(O_{1}^A, O_{1}^B), (O_{2}^A, O_{2}^B), \dots, $
where $O^A_k$ and $O^B_k$ are the number of events witnessed in each group in
the time interval $(t_{k - 1}, t_{k}]$, and $O_k = O^A_k + O^B_k$ is the total.
Notice that since the observation times are discrete, we can index the
observations by $k$ instead of $t_k$. For each $k$, let
$\bar{y}^A_{k} = \sum_{j\in A} y^j_{t_{k}}$ the number of participants at risk at
time $t_k$, define similarly $\bar{y}^B_{k}$, and let
$\bar{y}_k = \bar{y}^A_{k} + \bar{y}^B_{k}$ be the total. We derive an
anytime-valid test---a test valid at any observation time---for the problem
\eqref{eq:general_testing_problem}, where the hazard ratio under the null
hypothesis is $\theta_0 = 1$. The reason for this restriction in the null
hypothesis---only $\theta_0=1$ is allowed---will soon become clear. Observe
that, at time $t_k$, conditionally on $(\bar{y}^A_{k - 1}, \bar{y}^B_{k - 1})$
and the total number of events $O_k$, the number of events $O^B_k$ in group $B$
follows a hypergeometric distribution. This implies that, conditionally on
$(\bar{y}_{k-1}^A, \bar{y}_{k-1}^B, O_{k})$, the conditional likelihood of
observing $O^B_{k}$ is
$p_{k}(O^B_k) = p_{\mathrm{Hyper}}( O^B_k ; \ \bar{y}_{k - 1}, \bar{y}_{k -
  1}^B, O_k )$, where $p_{\mathrm{Hyper}}$ is the probability mass function of a
hypergeometric random variable, that is,
\begin{equation*}
  p_{\mathrm{Hyper.}}( o^B ; \ \bar{y},  \bar{y}^B, o)
  =
  \frac{
    \binom{\bar{y}^B}{o^B}
    \binom{\bar{y} - \bar{y}^B}{o - o^B}
  }{
    \binom{\bar{y}}{o}
  }.
\end{equation*}
With this observation at hand, we can build, analogously to
\eqref{eq:definition_generalized_statistic} from the continuous-monitoring case,
anytime-valid tests based on partial likelihood ratios,
\begin{equation}
  S^q_{1,k}\label{eq:generalized_statistic_ties}
  =
  \prod_{l\leq k}
  \frac{
    q_{l}( O^B_l)
  }{
    p_{l}( O^B_l)
  },
\end{equation}
where each $q_k$ is a conditional distribution on the possible values of $O^B_k$
that only depends on the data up to time $t_{k-1}$. Following the same steps as
in Section~\ref{sec:safe-logrank-test-1}, a sequential test based on monitoring
whether $S^q_{1,k}$ crosses the threshold $1 / \alpha$ is also anytime valid at
level $\alpha$.
\begin{lemma}
\label{lem:ties_test_martingale}
  Let $t_\kappa \in\{t_1, t_2, \dots\}$ be an arbitrary random time. The test
  $\xi_{1, \kappa}^{q}$ given by
  $\xi_{1,\kappa}^q = \indicator{S^q_{1,\kappa} \geq 1 / \alpha}$, where
  $S_{1,\kappa}^q$ is as in \eqref{eq:generalized_statistic_ties}, has
  type-I error bounded by $\alpha$, that is,
  \begin{equation*}
    \mathbf{P}_0\{\xi_{1,\kappa}^q = 1 \} \leq \alpha,
  \end{equation*}
  under any distribution $\mathbf{P}_0$ such that the hazard ratio is
  $\theta = 1$.
\end{lemma}
Just as in the proof of Proposition~\ref{prop:anytime-valid}, this lemma is shown by
a combination of the martingale property of $S_{1, k}^{\theta_1}$ and
Doob's maximal inequality. Therefore, we omit the proof of
Lemma~\ref{lem:ties_test_martingale}.

In order to obtain an optimal test under a particular hazard ratio
$\theta_1$---an alternative hypothesis---, it is necessary to compute the
partial conditional likelihood for the data under the alternative of having
observed $O^B$ given $(\bar{y}_{k-1}^A, \bar{y}_{k-1}^B, \bar{N}_{k-1})$. This
conditional likelihood is given by Fisher's noncentral hypergeometric
distribution with parameter $\omega$. Unfortunately, $\omega$ depends on the
baseline hazard function $\lambda$, which is assumed to be unknown (see
Appendix~\ref{app:ties} for details). It is for this reason that we restrict the
null hypothesis to $\theta_0 =1$. Luckily, since the test based on
$S^q_{\theta_0,t}$ remains valid even if $q$ is only approximately correct, this
problem can be skirted. As also noted by \cite{mehrotra_relative_2001}, when the
times between observations are short, the parameter $\omega$ is well
approximated by $\theta_1$, the hazard ratio under the alternative
hypothesis---no knowledge of $\lambda$ is needed for the approximation. With
this in mind, we put forward the use of $S_{\theta_0,k}^{\theta_1}$
\begin{equation*}
  S^{\theta_1}_{1,k}
  :=
  \prod_{l\leq k}
  \frac{
    p_{\theta_1, k}(O_{k}^B)
  }{
    p_{1, k}(O_{k}^B)
  }
\end{equation*}
where $S_{1,k}^{\theta_1}$ is a an instance
of~\eqref{eq:generalized_statistic_ties} with
$q_k(O^B_k) = p_{\theta_1, k}(O^B_k) $ and $p_{\theta_1, k}(O^B_k)$ is Fisher's
noncentral hypergeometric distribution with parameter $\omega = \theta_1$, that
is,
\begin{align*}
  p_{\theta_1, k}(o^B)
  &=
    p_{\mathrm{FNCH}}( o^B ; \ \bar{y},
    \bar{y}^B, o, \omega = \theta_1)\\
  &=
    \frac{
    \binom{\bar{y}^B}{o^B}
    \binom{\bar{y} - \bar{y}^B}{o - o^B}
    \theta_1^{o^B}
    }{
    \sum_{\max\{0, o^B - \bar{y}^B\}\leq u \leq \min\{ \bar{y}^B,  o^B\}}
    \binom{\bar{y}^B}{u}
    \binom{\bar{y} - \bar{y}^B}{o^B - u}
    \theta_1^{u}
    }.
\end{align*}
We remark that despite $p_{(\theta_1),k}$ being only approximately the correct
distribution for the observations under the alternative, type-I error guarantees
are not compromised (see the discussion on luckiness in
Section~\ref{sec:impr-altern}). In any case, this approximation is accurate when
the time between two consecutive observation times is not very long and when the
number of tied observations is small. Two reassuring remarks are in order. First,
in the special case when only one observation is made in each time interval
between two consecutive observation moments, the statistic $S_{1,k}^{\theta_1}$
reduces to the continuously monitored AV logrank test
(\ref{eq:def_safe_logrank_statistic}) at time $t_k$. Second, the score test
associated to $S_{1,k}^{\theta_1}$ coincides with the logrank test as is
conventionally computed in the presence of ties.

\section{A Gaussian approximation to the AV logrank test}
\label{sec:appr-safel-test}
In this section we present an approximation to the AV logrank test introduced in
the previous section. This is based on a sequential-Gaussian approximation to
the logrank statistic. The approximation is of interest for two reasons. First,
in practical situations, only the logrank $Z$-statistic (a standardized form of
the classic logrank statistic) and other summary statistics may be
available---and not the full risk-set process. This is often the case in medical
trials, where the full data sets are confidential. If we also know the number of
events $\bar{N}_k$ and the initial number of participants in both groups, $m^A$
and $m^B$, the Gaussian approximation to the AV logrank statistic can still be
used. The second reason, which we address in
Section~\ref{sec:rejection-regions}, is related to the fact that
$\alpha$-spending and group-sequential approaches, which we use as benchmarks,
are also based on Gaussian approximations to the classic logrank statistic.
Consequently, the behavior of the Gaussian approximation gives further insights
into how the AV logrank statistic compares to group-sequential and
$\alpha$-spending approaches as well. We henceforth focus on the main case of
interest, $\theta_0 = 1$.

Our general strategy is close in spirit to that followed in the construction of
the exact AV logrank statistic in Section~\ref{sec:safe-logrank-test-1}. We
build likelihood ratios using a classic approximation for the distribution of the
original logrank statistic \citep{schoenfeld_asymptotic_1981}. If the
distribution of this statistic was exactly normal, we could monitor continuously
its likelihood ratio. We show through extensive simulation in which regimes this
approximation behaves similarly to the AV logrank statistic.

We begin by recalling the definition of the $Z$-score associated to the classic
logrank test. Let $E^B_{i} = O_i p^B_{i}$ with
$p^B_{i} = \bar{y}^B_{i} / (\bar{y}^A_{i} + \bar{y}^B_{i})$ be the expected (under the null)
number of events witnessed in the time interval $(t_{i-1},t_{i}]$ in group $B$,
and let
$V^B_i = O_i \ p^B_{i}(1 - p^B_{i})\frac{\bar{y}_{i} - O_{i}}{\bar{y}_{i} - 1}$ be its variance.
After $k$ observations the $Z$-score associated to the classic
logrank statistic, $Z_k$, is given by
\begin{equation}
\label{eq:logrank_z_score}
  Z_{k} = \frac{\sum_{i \leq k} \bracks{O^B_{i} - E^B_{i}}}{\sqrt{\sum_{i\leq k}V_{i}^B}}.
\end{equation}
The numerator in the definition of $Z_{k}$ is the classic logrank statistic
$H_k = \sum_{i \leq k} \bracks{O^B_{i} - E^B_{i}}$, which is typically
interpreted as the cumulative difference between observed counts $O_{i}^B$ and
the expected counts $E_{i}^B$ in Group $B$. The factor
$\frac{\bar{y}_{i} - O_{i}}{\bar{y}_{i} - 1}$ found in $V^B_{i}$ can be
interpreted as a multiplicity correction, that is, a correction for ties
\citep[p. 207]{klein_survival_2003}. When only one event is witnessed between
two consecutive observation times, then $O_{i} = 1$, $E^B_{i} = p_{i}^B$, and
$V_i^B = p_i^B(1 - p_i^B)$. We remark that the above formulation is also found
in the work of \citet[(26)]{cox_regression_1972}.

We put forward the Gaussian approximation $S^{G}_{k}$ to the logrank statistic
$S^{\theta_1}_{1, k }$---we show its derivation in
Appendix~\ref{app:gaussian_details}---, given by
\begin{equation}
\label{eq:gaussian_safelogrank}
  S_{k}^{G}
  :=
  \exp\paren{
    -\frac12 \bar{N}_k\mu_1^2 + \sqrt{\bar{N}_k}\mu_1 Z_k},
\end{equation}
where $\bar{N}_k$ is the total number of observations up until time $t_k$ and
\begin{equation*}
  \mu_1 = \log(\theta)\sqrt{m^Bm^A / (m^A + m^B)^2}.
\end{equation*}
For an arbitrary random observation time $t_K \in \{t_1, t_2, \dots\}$, we refer
to the test $\xi^G_K = \indicator{S_K^G \geq 1 / \alpha}$ as the Gaussian AV
logrank test for~\eqref{eq:general_testing_problem}. Recall that we test
$\theta_0 = 1$, which corresponds to the asymptotic mean of the $Z$-score under
the null hypothesis being $\mu_0 = 0$. In Appendix~\ref{app:safety-only-balanced} extensive simulations
are performed to show in which regimes the Gaussian logrank test
retains type-I error guarantees. In Appendix~\ref{app:sample-size}, it is shown that,
under continuous monitoring, the Gaussian AV logrank test tends to be more
conservative---it needs more data than the exact one. The conclusion is the
following: \( S_{K}^G \) can be used for designs with balanced
allocation, and it approximates \( S_{1, K}^{\theta_{1}} \) well for hazard ratios between $0.5$
and $2$. 

We now compare the rejection regions defined by the Gaussian logrank test to
those of continuously monitoring using $\alpha$-spending and group-sequential
approaches.

\subsection{Rejection region and $\alpha$-spending}
\label{sec:rejection-regions}
In this section we compare the rejection regions of the $Z$-scores for which
$\alpha$-spending approaches and the AV logrank test for the null hypothesis
of no effect (hazard ratio $\theta_0 = 1$). The two main
$\alpha$-spending approaches discussed here are due to \citet{pocock_group_1977}
and \citet{obrien_multiple_1979}. We provide two reasons why the main focus of the
comparison, however, will be on the O'Brien-Fleming approach. Firstly, in retrospect,
Pocock himself believes that his approach leads to boundaries that are unsuitable \citep{pocock_current_2006}. One main feature of the Pocock procedure is that the
rejection regions are the same regardless of whether the (interim) analyses are
conducted at the start or the end of the trial. In practice this leads to many stopped
trials for benefits based on (too) small sample sizes and with unrealistically large
treatments effects \citep{pocock_current_2006}. In contrast, the rejection boundary
of the O'Brien-Fleming is more conservative at the start than at the end of the trial.
Secondly, the Pocock procedure only allows for a finite number of planned analyses
and, therefore, cannot be monitored continuously, whereas this is possible with
the O'Brien-Fleming $\alpha$-spending approach. Hence, the fair comparison is
between the two procedures (the AV logrank test and the O'Brien-Fleming
$\alpha$-spending approach) that allow for continuous monitoring.

We begin by specifying the rejection regions for both the Gaussian AV logrank
test and that of the O'Brien-Fleming $\alpha$-spending procedure. For the
Gaussian AV logrank we compute the region for the $Z$-score that rejects the
null hypothesis. Indeed, using \eqref{eq:gaussian_safelogrank}, we can compute
that whenever $m^A = m^B$, the null hypothesis is rejected as soon as
\begin{align*}
  Z_n
  \geq
  \frac{\sqrt{n}}{4} \ln (\theta_1)
  -
  \frac{2}{\sqrt{n}}\frac{
  \log(\alpha)
  }{
  \log (\theta_1)
  }
  & \text{ \ \ if  \ \ } \theta_1 > 1,\text{ or} \\
  Z_n
  \leq
  \frac{\sqrt{n}}{4} \ln (\theta_1)
  -
  \frac{2}{\sqrt{n}}
  \frac{
  \log(\alpha)
  }{
  \log (\theta_1)
  }
  & \text{ \ \ if \ \  } \theta_1 < 1.
\end{align*}

The O'Brien-Fleming procedure is based on a Brownian-motion approximation to
the sequentially computed logrank statistic $Z$-score. Indeed, for large values
of $n_{\max}$ and $t \in [0,1]$, the process
$t\mapsto\sqrt{\frac{\lfloor tn_{\max}\rfloor}{n_{\max}}} Z_{\lfloor t n_{\max}
  \rfloor}$ can be approximated by a Brownian motion $B_t$. We stress the fact
that $n_{\max}$ has to be set in advance. If $B_t$ is a Brownian motion, the
reflection principle, a well-known but nontrivial application of the symmetry of
$B_t$, implies that
\begin{equation*}
  \mathbf{P}\{ \max_{ 0 \leq t\leq 1} B_t \geq c\}
  =
  2\mathbf{P}\{B_1 \geq c\}
\end{equation*}
Since $B_1$ is Gaussian with mean zero and standard deviation $1$, setting
$c = q_{1 - \alpha / 2}$, the $(1 - \alpha/2)$-quantile of a standard Gaussian
distribution, then
\begin{equation*}
  \mathbf{P}\{ \max_{ 0 \leq t\leq 1} B_t \geq q_{1 - \alpha/2}\} =  \alpha.
\end{equation*}
This implies that
\begin{equation*}
  \mathbf{P}\{\max_{n\leq n_{\max}} \sqrt{n} Z_{n} \geq \sqrt{n_{\max}}q_{1 - \alpha/2}\}
  \approx \alpha,
\end{equation*}
or, in other words, the procedure that continuously monitors whether the
$Z$-score crosses the boundary $\sqrt{n_{\max}}q_{1 - \alpha/2}$ guarantees
approximate type-I error $\alpha$. Given a hazard ratio $\theta_1$ under the
alternative hypothesis, $n_{\max}$ can be set to achieve a desired type-II
error. The left-handed procedure can be worked out similarly, and we obtain
that, for $m^A=m^B$, the continuous-monitoring version of the O'Brien-Fleming
procedure rejects as soon as
\begin{align*}
  Z_n \geq \sqrt{\frac{n}{n_{\max}}} q_{1 - \alpha/2}
  &\text{ \ \ if \ \ } \theta_1 > 1  \text{ (right-sided test)}, \text{ or }\\
  Z_n \leq \sqrt{\frac{n}{n_{\max}}} q_{1 - \alpha/2}
  &\text{ \ \ if \ \ } \theta_1 < 1 \text{ (left-sided test)}.
\end{align*}

\begin{figure*}[h!]
  \centering
  \includegraphics[scale = 1]{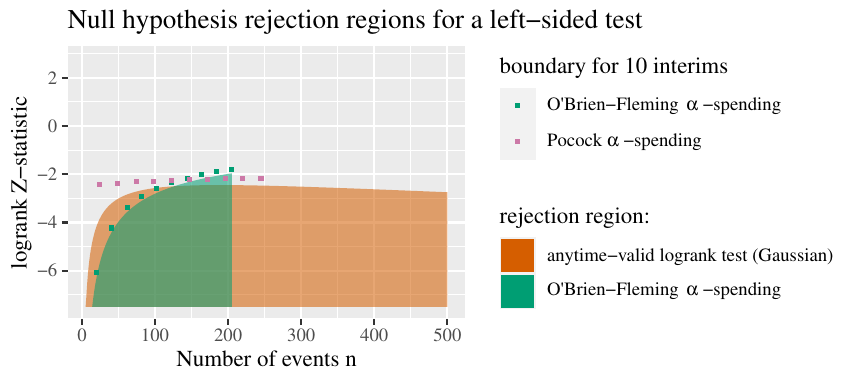}
  \caption{Left-sided rejection regions for continuous-monitoring using
    O'Brien-Fleming $\alpha$-spending or the Gaussian AV logrank test.
    Allocation is balanced ($m^A = m^B$) and $\alpha = 0.05$. Also shown are the
    O'Brien-Fleming and Pocock $\alpha$-spending boundaries for 10 interim
    analyses. The $\alpha$-spending boundaries are designed to have $80\%$ power
    when detecting a hazard ratio $0.7$. For more details, including the values
    of $n_{\max}$, see Section~\ref{sec:rejection-regions}. 
    \label{fig:rejectionRegions1}}
  \vspace{8mm} \includegraphics[scale = 1]{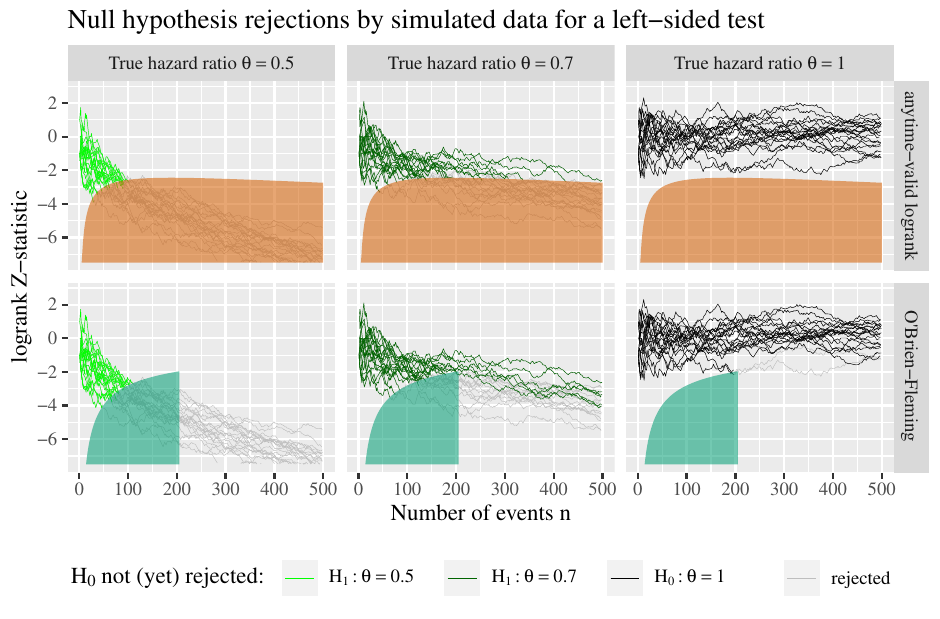}
  \caption{Null hypothesis rejections on simulated data. The rejection regions
    are the same as shown in \autoref{fig:rejectionRegions1} (designed to detect
    a hazard ratio of $0.7$ with $80\%$ power). Data are simulated under
    balanced allocation ($m_1=m_0 = 5000$) and as time-to-event data with
    possible ties. The logrank $Z$-statistic does not have a value for all $n$;
    it sometimes jumps with several additional events at a
    time. \label{fig:rejectionRegions2}}
\end{figure*}

The two regions of the $Z$-statistic values share an important feature: they are
more conservative to reject the null hypothesis at small sample sizes than at
larger ones, requiring more extreme values for the $Z$-statistic at the start of
the trial. This sets them apart from the Pocock spending function that requires
equally extreme values for the $Z$-statistic at small and large sample size.
\autoref{fig:rejectionRegions1} shows both the Gaussian AV logrank and the
O'Brien-Fleming $\alpha$-spending rejection regions. Additionally,
\autoref{fig:rejectionRegions1} shows the boundary of the Pocock
$\alpha$-spending function for 10 interim analyses. Note that the definition of
the AV logrank test rejection region requires a very explicit value for the
effect size $\theta_1 = \theta_{\min}$ of minimum clinical relevance, while that
value is implicit in the definition of the $\alpha$-spending rejection region:
To specify an maximum sample size $n_{\max}$ to achieve a certain power, an
effect size of minimal interest is also assumed. A fixed-sample-size analysis
designed to detect a minimum hazard ratio of $0.7$ would need 195 events to
achieve $80\%$ power if the true hazard ratio is also $0.7$. A sequential
analysis using $\alpha$-spending requires a slightly larger maximum number of
events: $205$ with the O'Brien-Fleming spending function; $245$, with the Pocock
$\alpha$-spending function---when we design for 10 interim analyses. We
investigate the number of events needed by the Gaussian AV logrank test in
Appendix~\ref{app:sample-size}. For the \( \alpha \)-spending procedures
continuing beyond \( n_{\max} \) is problematic. This is not the case for the AV
logrank test, as it allows for unlimited monitoring, then $n_{\max}$ is only a
soft constraint on the study---there is no penalty in type-I error for
continuing after $n_{\max}$ events have been witnessed.

The benefit of a sequential approach is that if there is evidence that the
hazard ratio is more extreme than it was anticipated under the alternative
hypothesis, we can detect that with fewer events than the maximum sample size. The left column of \autoref{fig:rejectionRegions2} illustrates that we benefit
because the true hazard ratio could be more extreme than we designed for (e.g.
$0.5$ instead of $0.7$; a larger risk reduction in the treatment group) and the
data reflects that. We also benefit from a sequential analysis if the true
hazard ratio is $0.7$ but by chance the values of our $Z$-statistics are more
extreme than expected. The major difference between $\alpha$-spending approaches
and the AV logrank test is that the AV test does not require to set a
maximum sample size. It in fact allows to indefinitely increase the sample size
without ever spending all $\alpha$. An $\alpha$-spending approach designed
to have $80\%$ power will miss out on rejecting the null hypothesis in $20\%$
(the type-II error) of the cases as is illustrate in the bottom middle plot of \autoref{fig:rejectionRegions2}
by the sample paths that remain (dark) green. In contrast, the AV logrank test
can potentially reject with \( 100\% \) power by continue sampling. In the sample
paths of $500$ events in \autoref{fig:rejectionRegions2}, all
but one sample path of $Z$-statistics could be rejected at a larger sample size by
the AV logrank test. By extending the trial, the AV logrank test can potentially
have $100\%$ power if the true hazard ratio is at least as small as the hazard
ratio set for minimum clinical relevance in the design of the test. Still,
type-I error is controlled. The bottom right plot of \autoref{fig:rejectionRegions2} shows two null
sample paths with a true hazard ratio of $1$ that are rejected by
the O'Brien-Fleming $\alpha$-spending region, but not by the AV logrank test.
Here, the AV logrank test is more conservative.

It is known that $\alpha$-spending methods behave poorly in case of unbalanced
allocation \citep{wu_group_2017}. In Appendix~{\ref{app:safety-only-balanced}}
we showed that our Gaussian approximation to the logrank test is also not an
$E$-variable in case of unbalanced allocation. Our exact AV logrank test, however,
is an $E$-variable under any allocation since it is defined directly on the risk-set process
\eqref{eq:one-outcome-evariable}. This suggests that if the complete data set is
available and allocation is unbalanced, the exact logrank test should be
preferred over the Gaussian approximation and the $\alpha$-spending methods.

\section{Optional continuation and live meta-analysis}
\label{sec:optional-continuation}
In this section, we address optional continuation and live meta-analysis---the
continuous aggregation of evidence from multiple experiments. For instance, data
could come from medical trials conducted in different hospitals or in different
countries. In such cases, we compare a global null hypothesis ${\cal H}_0$ that
is addressed in all trials (for instance, $\theta_0 = 1$) to an alternative
hypothesis ${\cal H}_1$ that allows for different hazard ratios in each
experiment. The present approach covers even the case in which the decision to
start each experiment might depend on the observations made in experiments that
are already in progress. Assume that there are $k_E$ experiments,
$E_{(1)}, \dots, E_{(k_E)}$, ordered by their respective starting times
$V_{(1)}\leq \dots\leq V_{(k_E)}$, each performed on different and independent
populations. Assume further that the starting time $V_{(k)}$, of experiment
$E_{(k)}$ depends only on the data observed in the ongoing experiments
$E_{(1)}, \dots, E_{(k-1)}$. If each experiment $E_{(k)}$ monitors the AV
logrank statistic $S^{k}_{\theta_0, t}$, where $S^{k}_{\theta_0,t} = 1$ for
$t \leq V_{(k)}$, then the product statistic
$S_{\theta_0,t}^{\mathrm{meta}}= \prod_{i\leq k_E} S^{i}_{\theta_0,t}$ is a test
martingale with respect to the filtration generated by all observations.
Consequently, the meta-test based on it enjoys anytime validity.
\begin{proposition}
\label{prop:meta-logrank-is-safe}
  Let $\tau$ be any random time. The test $\xi_{\theta_0, \tau}^{\mathrm{meta}}$
  given by $\indicator{S^{\mathrm{meta}}_{\theta_0,\tau} \geq 1 / \alpha}$,
  where $S^{\mathrm{meta}}_{\theta_0,t}= \prod_{i\leq k_E} S^{i}_{\theta_0,t}$,
  has type-I error smaller than $\alpha$.
\end{proposition}

This result follows from a reduction to independent left-truncation---we refer
to left-truncation in the specific sense defined by
\citet{andersen_statistical_1993}. Indeed, even in the presence of dependencies
on other studies, the observations made in $E_{(k)}$ can be regarded as a
left-truncated sample. Here, the time at which observation in $E_{(k)}$ is
started is random and only participants that have not witnessed an event are
recruited into the study. One may worry that these dependencies may alter the
sequential properties of $S^{\mathrm{meta}}_{\theta_0,t}$, but this is not the
case. Since the truncation time for $E_{(k)}$ is based on data that are
independent of that of experiment $E_{(k)}$---it is possibly based on the
observations made in all other experiments, it follows from results of
\citet{andersen_statistical_1993} (see Appendix~\ref{app:anytime-validity}) that
the sequential-multinomial interpretation of the partial likelihood for the
truncated data remains valid. Consequently, so does the sequentially computed AV
logrank statistic and the product statistic $S^{\mathrm{meta}}_{\theta_0,t}$. By
continuously monitoring $S^{\mathrm{meta}}_{\theta_0,t}$, we effectively perform
an \emph{online, cumulative} and possibly \emph{live} meta-analysis that remains
valid irrespective of the order in which the events of the different trials are
observed. Importantly, unlike in $\alpha$-spending approaches, the maximum
number of trials and the maximum sample size (number of events) per trial do not
have to be fixed in advance; we can always decide to start a new trial, or to
postpone to end a trial and wait for additional events.

\section{Anytime-valid confidence sequences}
\label{sec:confidence-sequences}
Anytime-valid (AV) confidence sequences corresponds to anytime-valid
tests in the same way fixed-sample tests correspond to confidence intervals.
Indeed, it is possible to ``invert'' a fixed-sample test to build a confidence interval:
the parameters of the null hypothesis that are not rejected by a the test form
a confidence interval. Analogously, test martingales can be used to derive AV
confidence sequences
\citep{darling_confidence_1967,lai_confidence_1976,howard_exponential_2018,howard_uniform_2018}.
In our setting, a $(1-\alpha)$-AV confidence sequence is a sequence of
confidence intervals $\{\mathrm{CI}_t\} _{t\geq 0}$, such that
\begin{equation}
\label{eq:ci}
  \mathbf{P}_{\theta}\{
  \theta \notin  \mathrm{CI}_{t}  \  \text{ for some } \  t\geq 0
  \}\leq \alpha.
\end{equation}
A standard way to design $(1-\alpha)$-AV confidence sequences, translated to our
logrank setting, is to use a prequential plug-in test martingale
$S^{\mathrm{preq}}_{\theta_0,t}$ or the Bayesian version
$S^{\mathbf{W}}_{\theta_0,t}$ as in Section~\ref{sec:impr-altern}. At time $t$,
one reports $\mathrm{CI}_{t} = [\theta_{t}^L, \theta_{t}^U]$ where
$\mathrm{CI}_t$ is the smallest interval containing the values of $\theta_0$
such that $S_{\theta_0,t}^{\mathrm{preq}} > 1/\alpha$ outside this interval.
Ville's inequality readily implies that this is indeed an AV confidence
sequence. The same construction can be made for arbitrary instances of
$S^q_{\theta_0, t}$ as in (\ref{eq:definition_generalized_statistic}).

\section{Power and sample size}
\label{sec:power-sample-size}
\begin{figure*}[h!]
  \centering
  \includegraphics[scale = .97]{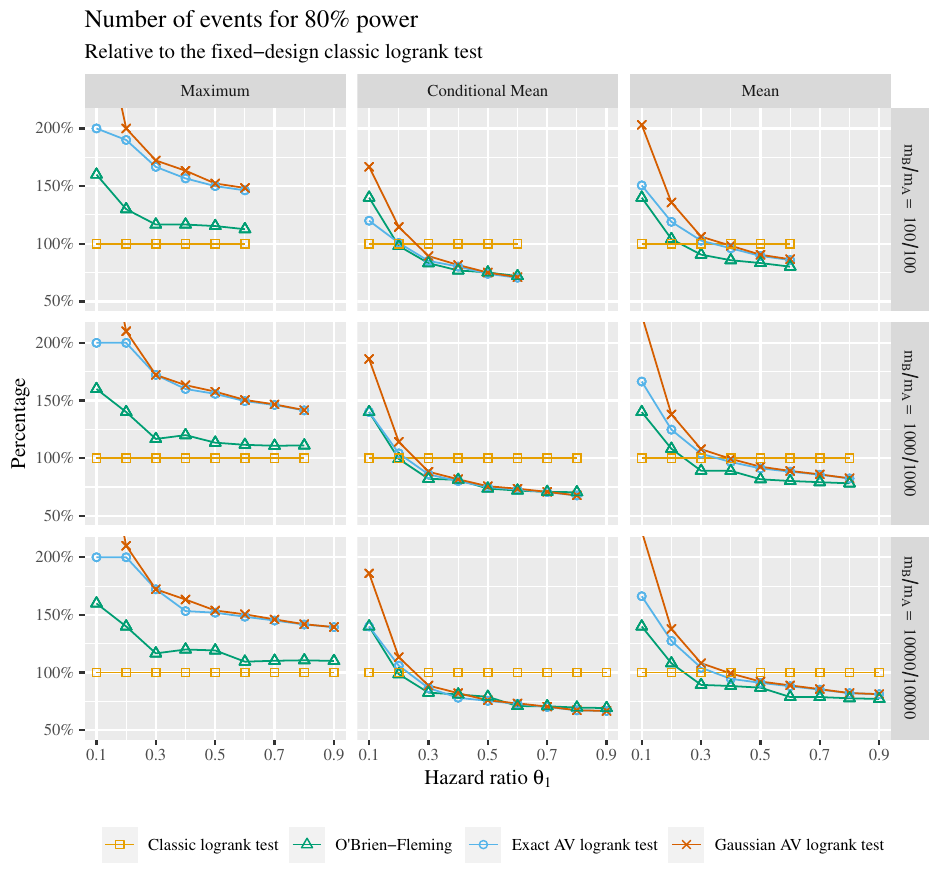}
  \caption{Maximum, expected (Mean) number of events needed to reject the null
    hypothesis with $80\%$ power. `Conditional Mean' makes reference to the
    number of events needed given that the null hypothesis is indeed rejected.
    The maximum number of events needed using AV logrank statistics is higher
    than that of a fixed-sample test, but lower in expectation (see
    Section~\ref{sec:power-sample-size}). All simulations are performed with
    $\alpha = 0.05$ and tests are designed to detect the hazard ratio $\theta_1$
    shown on the x-axis. Data are generated using that same hazard ratio. The
    classical logrank test needs the following sample sizes (number of events)
    $n(\theta_1)$ for an $80$\%-power design to detect hazard ratio $\theta_1$:
    $n(0.1) = 5, \, n(0.2) = 10, \, n(0.3) = 18, n(0.4) = 30, \, n(0.5) = 52, \,
    n(0.6) = 95, \, n(0.7) = 195, \, n(0.8) = 497 \text{ and } n(0.9) = 2228$.
    These sample sizes represent the $100\%$ line in all plots.
    \label{fig:simulation_expected_vs_worst} }
\end{figure*}

In this section, we investigate the power properties of the AV logrank test---we
will study specific stopping times.
%
%
We have seen that by observing arbitrarily long sequences of events the logrank
test can achieve type-II errors that are as close to zero as desired. However,
in practice it is necessary to plan for a  maximum number of events
$n_{\max}$ so that either the experiment is stopped as soon as the null
hypothesis is rejected or when $n_{\max}$ events have been observed. In the
latter case, there is no evidence to reject the null hypothesis. We assess via
simulation the value of $n_{\max}$ needed to guarantee $20\%$ type-II error
($80\%$ power) for the exact and Gaussian AV logrank tests. We compare this to
the $n_{\max}$ needed to achieve the same power using the continuous-monitoring
O'Brien-Fleming $\alpha$-spending procedure introduced in the previous section,
and the fixed-sample-size classic logrank test.
Figure~\ref{fig:simulation_expected_vs_worst} show simulation results
establishing three types of sample sizes. The leftmost panels (``Maximum'')
shows the sample size $n_{\max}$ described earlier, which would be required to
design the experiment. We stress the fact that using the classic logrank test or
$\alpha$-spending designs events beyond $n_{\max}$ cannot be analyzed. The
rightmost panel of Figure~\ref{fig:simulation_expected_vs_worst} (``Mean'')
shows the sample sizes that capture the expected duration of the trial. It
expresses the mean number of events, under the alternative hypothesis, that will
be observed before the trial can be stopped. Here, for the AV logrank tests, we
use the aggressive stopping rule that stops as soon as
$S_{\theta_0, t}^{\theta_1}\geq 1/\alpha = 20$ or $n = n_{\max}$. In case of
$\alpha$-spending approaches and the AV logrank test this number of events is
always smaller than the maximum needed in the design stage. Lastly, the middle
panel (``Conditional Mean'') shows an even smaller number for those tests that
have a flexible sample size: the expected stopping time {\em given\/} that the
trial is stopped before the maximum $n_{\max}$ was reached---this only happens
if the null is rejected. For comparison purposes, all sample sizes are shown
relative to (i.e., divided by) the fixed sample size needed by the classical
logrank test to obtain $80\%$ power. Note that for small sample size (for small
hazard ratios), both the classic logrank test and O'Brien-Fleming
$\alpha$-spending are not recommended due to lack of type-I error control. They
are based on Schoenfeld's Gaussian approximation, which underestimates the
number of events required for hazard ratios far away from 1. For example,
simulations show that for $\theta_1 = 0.1$, $n = 6$ or $7$ events will be
necessary---for small sample sizes the classical logrank test is not recommended
due to lack of type-I error control. We give further details in
Appendix~\ref{app:moreSampleSize} (see also
Figure~\ref{fig:simulation_expected_vs_worst}). In summary, at all hazard ratios
at which the Gaussian approximation to the classic logrank test is accurate (say
for $\theta_1 \geq 0.3$), the mean number of events needed by the AV logrank
tests is about the same or noticeably smaller than that needed when using a
fixed-sample-size analysis.

\section{Discussion, Conclusion and Future Work}
\label{sec:disc-concl-future}
We introduced the AV logrank test, a version of the logrank test that
retains type-I error guarantees under optional stopping and continuation.
Extensive simulations reveal that, if we do engage in optional stopping, it is
competitive with the classic logrank test (which neither allows in-trial
optional stopping nor optional continuation) and $\alpha$-spending procedures (which allows
forms of optional stopping but not optional continuation). We provided an
approximate test for applications in which only summary statistics are available
and also showed how the AV logrank test can be used in combination with
(informative) priors and prequential learning approaches, when no effect size of
minimal clinical relevance can be specified. Two of our extensions invite
further research: we introduced anytime-valid confidence sequences for the
hazard ratio, and will study their performance in comparison to other approaches
in future work. We also introduced an extension to Cox' proportional hazards
regression, which guarantees type-I error guarantees even if the alternative
model is equipped with arbitrary priors. In future work, we plan to implement
this extension---which requires the use of sophisticated methods for estimating
mixture models. The GROW AV logrank tests (exact and Gaussian) are already
available in our \texttt{safestats} R package \citep{turner_safestats_2022}. We end with
two final points of discussion: {\em staggered entries\/} and {\em doomed
  trials\/}.

\subsection{Staggered entry} Earlier approaches to sequential time-to-event
analysis were also studied under scenarios of staggered entry, where each
patient has its own event time (e.g., time to death since surgery), but patients
do not enter the follow-up simultaneously (such that the risk set of, say, a
two-day-after-surgery event changes when new participants enter and survive two
days). \citet{sellke_sequential_1983} and \citet{slud_sequential_1984} show
that, in general, martingale properties cannot be preserved under such staggered
entry settings, but that asymptotic results are hopeful \citep{sellke_sequential_1983} as
long as certain scenarios are excluded \citep{slud_sequential_1984}. When all
participants' risk is on the same (calendar) time scale (e.g., infection risk in
a pandemic; staggered entry now amounts to left-truncation, which we can deal
with), or new patients enter in large groups (allowing us to stratify),
staggered entry poses no problem for our methods. But research is still ongoing
into those scenarios in which our inference is fully AV for patient time under
staggered entry, and those that need extra care.

\subsection{Your trial is not doomed} In their summary of conditional power
approaches in sequential analysis \citet*{proschan_statistical_2006} write that
low conditional power makes a trial futile. Continuing a trial in such case
could only be worth the effort to rule out an effect of clinical relevance, when
the effect can be estimated with enough precision. However, if \enquote{both
  conditional and revised unconditional power are low, the trial is doomed
  because a null result is both likely and uninformative} \citep[p.
63]{proschan_statistical_2006}. While this is the case for all existing
sequential approaches that set a maximum sample size, this is not the case for
AV tests. Any trial can be extended and possibly achieve 100\% power or in an
anytime-valid confidence sequence show that the effect is too small to be of
interest. This is especially useful for time-to-event data when sample size can
increase by extending the follow-up time of the trial, without recruiting more
participants. Moreover, new participants can always be enrolled either within
the same trial or by spurring new trials that can be combined indefinitely in a
cumulative meta-analysis.

\section*{Acknowledgements}
This work is part of the research program with project number 617.001.651, which is  financed by the Dutch Research Council (NWO). We thank Henri van Werkhoven, Richard Gill, Wouter Koolen, Aaditya Ramdas and Rosanne Turner for useful conversations.

\DeclareRobustCommand{\VOORVOEGSEL}[3]{#3}

\bibliographystyle{plainnat}
\bibliography{phd_basic, references_Judith}

\pagebreak

\appendix

\section{Omitted Proofs and Details}
\label{sec:math}
In this section we provide proofs and remarks omitted from previous sections. In
Appendix~\ref{app:math-optimality} we relate growth-rate
optimality to the minimum expected stopping time. In
Appendix~\ref{app:anytime-validity}, we show that the AV logrank statistic is a
continuous-time martingale, and show that this is also true for general patterns of incomplete
observation, such as left truncation and filtering as a consequence of the results of
\citet{andersen_statistical_1993}. In Appendix~\ref{app:ties},
we proof the claims made in Section~\ref{sec:ties-1} about the martingale
structure of the AV logrank test under the presence of ties. Lastly, in
Appendix~\ref{app:moreSampleSize}, we give further details on the simulations
used to compute the planned maximum sample sizes for a given targeted power.
Under the alternative and optional stopping, the observed sample size is in many cases lower.

\subsection{Expected Stopping Time, GROW and Wald's Identity}
\label{app:math-optimality}
Here we motivate the GROW criterion by showing that it minimizes, in a
worst-case sense, the expected number of events needed before there is
sufficient evidence to stop. 
Let $\mathbf{P}_{0}$ represent our null model, and let, as before, the alternative
hypothesis be ${\cal H}_1: \theta \leq \theta_1 $ for some $\theta_{1} < \theta_{0}$.
Suppose we perform a level-$\alpha$ test based on a test martingale
$S_{\theta_0, t}^{q}$ using the stopping rule $\tau$ that stops as soon as
$S_{\theta_0, t}^{q} $ exceeds the threshold $ 1/\alpha$, that is,
$\tau^q = \inf_{t} \{ t \, : \, S_{\theta_0, t}^{q} \geq 1/\alpha \}$. In the main text
we elaborated on how $S_{\theta_0, t}^{\theta_1}$ is optimal with respect to
the GROW criterion. We now show that the problem of minimizing the worst-case,
the expected number of events $\mathbf{E}_{\theta}[\bar{N}_{\tau^q}]$ over $q$ is
approximately equivalent to finding the GROW test martingale. To do so,
we make simplifying assumptions that reduce the problem to an i.i.d. experiment.
This allows us to employ a standard argument based on an identity of
\citet{wald_sequential_1947}, originally due to \citet{breiman_optimal_1961}. For this
we assume that the initial risk sets (i.e., $\bar{y}^A_0$ and $\bar{y}^B_0$) are
large enough so that, for all sample sizes we will ever encounter,
$\bar{y}^A_t / \bar{y}^B_t \approx \bar{y}^A_0 / \bar{y}^B_0 $. This allows us
to treat the likelihood of the participant(s) $I_{(k)}$ having witnessed the event at
time $T^{(k)}$ to be independent of $t$, that is, as an i.i.d. experiment.

The argument of \citet{breiman_optimal_1961} relates the expected number of
events to the expected value of our stopped AV logrank statistic. Suppose
first that we happen to know that the data come from a specific $\theta$ in the
alternative hypothesis. Then $S_{\theta_0, \tau}^{q}$ is the product of
$\bar{N}_\tau$ factors of ratios $R_{\theta_0, (i)}^{q} = q_{(i)}(I_{(i)}) / p_{\theta_0,(i)}(I_{(i)})$ at the \( i \)th event. Wald's identity applied to its logarithm implies
\begin{equation}
\label{eq:wald_identity}
  \mathbf{E}_{\theta}[\bar{N}_\tau]
  =
  \frac{
    \mathbf{E}_{\theta}[\ln S_{\theta_0, \tau^q}^{q}]
  }{
    \mathbf{E}_{\theta}[\ln R_{\theta_0,(1)}^{q}].
  }.
\end{equation}
For simplicity we will further assume that the number of participants at risk is
large enough so that the probability that we run out of data before we can
reject is negligible. Because of the choice of the stopping rule $\tau^q$, the
right-hand side of the last display can then be further rewritten as %
\begin{equation*}
  \frac{
    \mathbf{E}_{\theta}[\ln S_{\theta_0, \tau^q}^{q}]
  }{
    \mathbf{E}_{\theta}[\ln R_{\theta_0, (1)}^{q}]
  }
  =
  \frac{
    \ln(1 / \alpha)
    +
    \text{\textsc{very small}}
  }{
    \mathbf{E}_{\theta}\sqbrack{\ln \paren{q_{(1)}(I_{(1)}) / p_{(1), \theta_0}(I_{(1)})}}
  },
\end{equation*}
%
%
where \textsc{very small} between $0$ and $\log |\theta_1/\theta_0|$. The
equality follows because we reject as soon as
$S_{\theta_0, t}^{q} \geq 1/\alpha$, so $S_{\theta_0, \tau}^{q}$
cannot be smaller than $1/\alpha$, and it cannot be larger by more than a factor
equal to the maximum likelihood ratio at a single outcome (if we would not
ignore the probability of stopping because we run out of data, there would be an
additional small term in the numerator).

With~(\ref{eq:wald_identity}) at hand, we can relate our choice of $q$ to the
expected number of events witnessed before stopping. If, for a fixed $\theta$,
we try find the $q$ that minimizes the expected number of events
$\mathbf{E}_\theta[\bar{N}_{\tau^q}]$, and, as is customary in sequential
analysis, we approximate the minimum by ignoring the \textsc{very small} part,
we see that the expression is minimized by maximizing the numerator
$\mathbf{E}_{\theta}\sqbrack{\ln \paren{Q_{(1)} / P_{\theta_0, (1)}}}$ over $q$.
The maximum is achieved by $Q_{(1)}= P_{\theta, (1)}$; the expression in the
denominator then becomes the Kulback-Leibler divergence between two Bernoulli
distributions. It follows that, under $\theta$, the expected number of outcomes
until rejection is minimized by $Q_{(1)} = P_{\theta}$. Thus, in this case, we
use the GROW $S^{\theta}_{\theta_0, t}$ as test statistic. However, we still
need to consider the fact that the real ${\cal H}_1$ is composite: as
statisticians, we do not know the actual $\theta$; we only know
$0 < \theta \leq \theta_1$. A worst-case approach uses the $q$ achieving
\begin{equation*}
  \max_{q} \min_{\theta \leq \theta_1}
  \mathbf{E}_{\theta}\sqbrack{\ln \paren{p_{(1)}(I_{(1)}) / q_{(1), \theta_0}(I_{(1)})}}
\end{equation*}
%
%
since, repeating the reasoning leading to (\ref{eq:wald_identity}), this
$q$ should be close to achieving the min-max number of events until
rejection, given by
\begin{equation*}
  \min_{q} \max_{\theta \leq \theta_1}
  \mathbf{E}_{\theta} [\bar{N}_{\tau^q}]
\end{equation*}
But this just tells us to use the GROW $\E$-variable relative to ${\cal H}_1$,
which is what we were arguing for.

\subsection{Continuous time and anytime validity}
\label{app:anytime-validity}
In this section, we show the anytime validity of the AV logrank test. This is
done via Ville's inequality for which it suffices to show that
\( S^{q}_{\theta_{0}} = (S^{q}_{\theta_{0}, t})_{t \geq 0} \) is a nonnegative
(super) martingale. To do so, we use the counting process formalism. A few
definitions are in order. Only in this section, we assume knowledge of counting
process theory \citep[see][]{andersen_statistical_1993, fleming_counting_2011}.
Denote, for $i = 1,\dots,m$, $\tilde{N}_t^i = \indicator{ t \leq T^i}$ the
counting processes associated to each participant, and let $y^i_t$ be the
at-risk process. For each participant, the censored process $N_t^i$, which is
observed, is given by $\rmd N^i_t = y^i_t\rmd \tilde{N}^i_t$---we use this
convention to signify that $N^i_t = \int_0^ty^i_s\rmd \tilde{N}^i_s$. We define
the sigma-algebra
\( {\cal F}_t := \sigma(N_s^j: 0\leq s \leq t, j = 1,\dots, n) \), which, as
usual, can be interpreted as the information in the study up to time $t$.

One of the results of the counting process theory is that the processes
$\rmd N_t^i - y_t^i \rmd \lambda_t^i$ are martingales, where, recall,
$y^i_t = \indicator{X^i\geq t}$ is the at-risk process, and $\lambda_t^i$ is the
hazard function associated to $T^i$. In that case, $y^i_t\rmd\lambda_t^i$ is
called the compensator of $N_t^i$. The result that the AV logrank test is a
martingale hinges specifically on this structure. Thus, any pattern that
preserves this martingale structure also preserves the martingale property for
the AV logrank test, and consequently its type-I error guarantees.
\citet[III.4]{andersen_statistical_1993} show exactly this under general
patterns of incomplete observation provided that the mechanisms are independent
of the observations. With this in mind, in the following, we only assume that
the counting processes $N^i_t$ have compensators $A^i_t$ given by
$\rmd A^i_t = y^i_t \rmd \lambda^i_t$.

The filtration ${\cal F} = ({\cal F}_s)_{s\geq 0}$ is right-continuous and we
can safely identify predictable processes with left-continuous process. For some
$\theta_0$, denote by $\mathbf{P}_0$ the distribution under which, for each
$i = 1,\dots, m$, the hazard function for $T^i$ is
$\lambda^i_t = \theta_0^{z^i}\lambda_t^{A}$, where \( g^i = 0 \) if
\( i \in A \) and \( g^{i} =1 \) if \( i \in B \). Recall from
Section~\ref{sec:model}, if participant $i$ belongs to Group $B$,
$\lambda^i_t = \theta_0\lambda^B_t = \theta_0 \lambda^{A}_t$; otherwise,
$\lambda_t^i = \lambda^A_t$. Let $q^1_t, \dots, q^m_t$ be predictable processes
such that $\sum_{i\leq m}q^i_t y^i_t = 1$ a.s. for all $t$, that is,
$\{q^i_t\}_{i \in \mathcal{R}_t}$ at each \( t \) is a probability distribution
over the participants at risk at time $t$. Define $r^i_t$ to be each of the
ratios $r^i_{t} = q^i_t / p^i_{\theta_0, t}$. Define the predictable process
$S_{\theta_0, t^-}^q = \lim_{s \uparrow t} S_{\theta_0,t^-}^{q}$. As such, at
each $t$, the change
$\rmd S_{\theta_0, t}^{q} = S_{\theta_0,t}^q - S_{\theta_0, t^-}^q$ of the AV
logrank statistic $S_{\theta_1}^q$ at time $t$, given
in~\eqref{eq:definition_generalized_statistic}, can be computed as
\begin{equation*}
  \rmd S_{\theta_0,t}^q
  =
  \sum_{i \leq m} S_{\theta_0,t^-}^q (r^i_t - 1)\rmd N^i_t,
\end{equation*}
%
%
because no two events happen simultaneously with positive probability. Since
$S_{\theta_0,t^-}^q$ is predictable, it is enough to prove that the process
$M_t$ defined by $\rmd M_t = \sum_{i \leq m} (1 - r^i_t) \rmd N_t^i$ is a
martingale \citep[see][Theorem 1.5.1]{fleming_counting_2011}. Recall that
$\bar{y}^A_t = \sum_{i \in A} y^i_t$ and $\bar{y}^B_t = \sum_{i \in B} y^i_t$.
Then both $\bar{y}^A$ and $\bar{y}^B$ are left-continuous processes.
%
%
\begin{lemma}
\label{lem:martingale-property}
Let $\{q^i_t\}_{i \leq m}$ be a collection of nonnegative left-continuous
processes $q^i = (q^i_t)_{t\geq 0}$ such that $\sum_{i \leq m} y^i_tq^i_t = 1$
for all $t$. Let $\{p^i_{\theta_0, t}\}_{i \leq m}$ be the collection of
processes given by
  \begin{equation*}
    p^i_{\theta_0, t}
    =
    \frac{\theta_0^{g^i}y^i_t}{\bar{y}_t^A + \theta_0\bar{y}^B_t }.
  \end{equation*}
%
%
%
  The process $M = (M_t)_{t\geq0}$ given by
  $\rmd M_t = \sum_{i \leq m}(1 - r^i_t)\rmd N_t^i$ is a martingale under
  $\mathbf{P}_{0}$ with respect to the filtration
  ${\cal F} = ({\cal F}_t)_{t\geq 0}$.
\end{lemma}
\begin{proof} 
  It suffices to show that the compensator $A_t$ of $M_t$, given by
  $\rmd A_t = \sum_{i\leq m } \sum_{i \leq m} (r^i_t - 1)y^i_t\lambda^i_t\rmd t$
  is zero. Define $\bar{q}^A_t = \sum_{i\in A}y^i_tq^i_t$ and
  $\bar{q}^B_t = \sum_{i\in B}y^i_tq^i_t$. Notice that by assumption
  $\bar{q}^A_t + \bar{q}^B_t = 1$., and recall that, under the null
  \( \lambda^{B}_{t} = \theta_{0} \lambda^{A}_{t} \). We can compute
  \begin{align*}
    \sum_{i \leq m} (r^i_t - 1)y^i_t\lambda^i_t
    &= \sum_{i\in A}y^i_t\lambda^A_t(r^i_t - 1)  + \sum_{i\in B}y^i_t\lambda^B_t(r^i_t - 1)\\
    &=\begin{multlined}
      \lambda^A_t
      [
      (\bar{y}^A_t + \theta_0\bar{y}^B_t) \bar{q}^A_t -  \bar{y}^{A}_t +
      \\
      (\bar{y}^A_t + \theta_0\bar{y}^B_t)\bar{q}^B_t - \theta_0\bar{y}^B_t
      ]
    \end{multlined}\\
    &=
      \begin{multlined}
        \lambda^A_t[(\bar{y}^A_t + \theta_0\bar{y}^B_t)\overbrace{(\bar{q}^A_t +
          \bar{q}^B_t)}^{ = 1} - \\ (\bar{y}^A_t + \theta \bar{y}^B_t)]
      \end{multlined}
    \\
    &= 0,
  \end{align*}
  where we used the assumption that
  $\sum_{i \leq m} y^i_t q^i_t = \bar{y}^{A}_tq^{A}_t + \bar{y}^{B}_tq^{B}_t =
  1$. As the compensator $A_t$ of $M_t$ is zero at each \( t \), we conclude
  that \( M_t \) is a martingale, as was to be shown.
\end{proof}

Our previous discussion and the preceding lemma have the following corollary as
a consequence.
\begin{corollary}
  $S_{\theta_0}^q = (S_{\theta_0,t}^q)_{t\geq 0}$ is a nonnegative martingale with
  expected value equal to one.
\end{corollary}
Hence, Ville's inequality holds for \( S_{\theta_0}^q \), which implies that
\begin{equation*}
  \mathbf{P}_0\{S_{\theta_0,t}^q \geq 1  / \alpha \text{ \ for some \ } t\geq 0\} \leq
  \alpha .
\end{equation*}
This implies the anytime validity of the test
$\xi_{\theta_0}^q = (\xi_{\theta_0,t}^q)_{t\geq 0}$ given by the AV logrank
test $\xi_{\theta_0, t}^q = \indicator{S_{\theta_0, t}^q \geq 1/ \alpha}$.

\subsection{Ties}
\label{app:ties}
The purpose of this section is twofold. Firstly, we prove
Lemma~\ref{lem:ties_test_martingale}. Secondly, we show that the conditional
likelihood given in Section~\ref{sec:ties-1} indeed approximates the true
conditional partial likelihood ratio under any distribution such that the hazard
ratio is $\theta_1$.
%
%

Our general strategy in this case is similar to the one undertaken in the
continuous-monitoring case: we build a test martingale with respect to a
filtration ${\cal G}^\star$, and use Ville's inequality to derive anytime-valid
type-I error guarantees. Define, for each $k = 1, 2, \dots$, the sigma-algebra
${\cal G}_k$ generated by all observations made in times $t_1, \dots, t_k$, that
is,
${\cal G}_{k} = \sigma(N_{t_l}^i, \tilde{N}_{t_l}^i \ : \ i = 1, \dots, m; \ l =
1,\dots, k)$, and the corresponding filtration
${\cal G} = ({\cal G}_k)_{k =1, 2, \dots}$. Under Cox's proportional hazard
model, conditionally on ${\cal G}_{k - 1}$, our observations
$\Delta \bar{N}^A_k$ and $\Delta \bar{N}^B_k$ are binomially distributed with
parameters depending on the hazard function (see
Lemma~\ref{lem:observation_counts_binomial} below). By conditioning both on
${\cal G}_{k-1}$ and on the total number of events
%
%
$\Delta \bar{N}_k = \Delta \bar{N}^A_k + \Delta \bar{N}^B_k$, we use the
likelihood of having observed $\Delta \bar{N}^B_k$, which follows Fisher's
noncentral hypergeometric distribution, as detailed in
Corollary~\ref{cor:conditional_countsB_hypergeometric}. We gather these
observations in the following two lemmas.
%
%
\begin{lemma}
\label{lem:observation_counts_binomial}
  Conditionally on ${\cal G}_{k-1}$, the following hold:
  \begin{enumerate}
  \item The number of events $\Delta \bar{N}^A_k$ has a binomial distribution
    with parameters $\bar{y}^A_k$ and $p^A_k$ where
    $p^A_k = 1 - \exp\paren{ - \int_{t_{k - 1}}^{t_k}\lambda_s^A \rmd s}$.
  \item The number of events $\Delta \bar{N}^B_k$ has a binomial distribution
    with parameters $\bar{y}^B_k$ and $p^B_k$ where
    $p^B_k = 1 - \exp\paren{ - \theta\int_{t_{k - 1}}^{t_k}\lambda_s^A \rmd s}$
    and $\theta$ is the hazard ratio.
  \end{enumerate}
\end{lemma}
\begin{proof}
  The result is standard, and it follows from explicitly solving for $\lambda$
  in \eqref{eq:definition_hazard_function} and computing the conditional
  probability in \eqref{eq:cox_conditional_jump_probability} for each group.
\end{proof}

Next, we use a standard result: given two binomially distributed random
variables $X$ and $Y$, the distribution of $X$ conditionally on $X + Y$ is
Fisher's noncentral hypergeometric distribution. We apply this to
$\Delta\bar{N}^A_k$ and $\Delta\bar{N}^B_k$ from the previous lemma and spell
out the corresponding parameters in the following corollary.

\begin{corollary}
\label{cor:conditional_countsB_hypergeometric}
  Let
  ${\cal G}^\star_{k - 1} = {\cal G}_{k - 1}\vee \sigma(\Delta \bar{N}_{k})$,
  and let $p_k^A$ and $p_k^B$ as in Lemma~\ref{lem:observation_counts_binomial}.
  Define the odd ratios $\omega^A_k = p^A_k / (1 - p^A_k)$,
  $\omega^B_k = p^B_k / (1 - p^B_k)$ and the ratio
  $\omega_k = \omega^B_k / \omega^A_k$. Then, conditionally on
  ${\cal G}^\star_{k - 1}$, the likelihood of having observed
  $\Delta \bar{N}^B_k$ events in group $B$ is given by Fisher's noncentral
  hypergeometric distribution
  $p_{\mathrm{FNCH}}(\Delta \bar{N}^B_k \ ; \ \bar{y}^B_{k - 1}, \bar{y}^A_{k -
    1}, \Delta\bar{N}_k, \omega_k)$, where
  \begin{equation*}
    \begin{multlined}
      p_{\mathrm{FNCH}}( n^B ; \ \bar{y}^B,
      \bar{y}^A, n, \omega)
      =\\
      \frac{
        \binom{\bar{y}^B}{n^B}
        \binom{\bar{y}^A}{n - n^B}
        \omega^{n^B}
      }{
        \sum_{\max\{0, n^B - \bar{y}^B\}\leq u \leq \min\{ \bar{y}^B,  n^B\}}
        \binom{\bar{y}^B}{u}
        \binom{\bar{y}^A}{n^B - u}
        \omega^{u}
      }.
    \end{multlined}
  \end{equation*}
\end{corollary}
Naively, one could use a partial likelihood ratio just as in the absence
of ties to derive a sequential test. This, however, is not satisfactory, because, in
general, the parameter $\omega_k$ depends heavily on the unknown baseline hazard function $\lambda^A$. Contrary to the general case, when the hazard ratio $\theta$ is one,
the parameter $\omega_k = 1$, and Fisher's noncentral hypergeometric
distribution reduces to the conventional hypergeometric distribution. With this
%
%
observation at hand, if $\{q_k\}_{k = 1, 2, \dots}$ is a sequence of conditional
distributions $q_k(\ \cdot \ )$ on the possible values of $\Delta\bar{N}^B_k$,
we can build a sequential tests for~\eqref{eq:general_testing_problem}, with its
corresponding type-I error guarantee. We give the details in the following
corollary, and subsequently point at a useful choice for $q$ that approximates
the real likelihood.

The choice of $q$ for our statistic presented in Section~\ref{sec:ties-1}
follows from an approximation of the parameter $\omega$ for small
$\Delta t_k = t_{k} - t_{k - 1}$. As noted by \cite{mehrotra_relative_2001}, if
$\int_{t_{k-1}}^{t_k}\lambda_1(s) \rmd s$ is small, then
$p^A_k\approx \lambda_{t_{k - 1}}\Delta t_k$ and $p^A_k \approx \theta p^A_k$.
With these two approximations, $\omega_k \approx \theta$. This means that the
choice
$q_k(\Delta \bar{N}^B_k) = p_{\theta_1, k}(\Delta \bar{N}^B_k):=
p_{\mathrm{FNCH}}(\Delta \bar{N}^B_k ; \ \bar{y}^B_k, \ \bar{y}^A_k, \
\Delta\bar{N}_k, \ \omega = \theta_1)$ approximates the real conditional
likelihood under any alternative for which the true hazard ratio is $\theta_1$.
Hence, the sequentially computed statistic
\begin{equation*}
  S^{\theta_1}_k
  =
  \prod_{l\leq k}
  \frac{
    p_{\theta_1, k}(\Delta \bar{N}_{k}^B)
  }{
    p_{1, k}(\Delta \bar{N}_{k}^B)
  }
\end{equation*}
approximates the true partial likelihood ratio of the data observed up to time
$t_k$ in the presence of ties, and we recommend its use.

\subsection{Details of sample size comparison simulations}
\label{app:moreSampleSize}
In this section we lay out the procedure that we used to estimate the expected
and maximum number of events required to achieve a predefined power as shown in
\autoref{fig:simulation_expected_vs_worst} and
\autoref{fig:simulation_sampleSize_misspecified} in
Section~\ref{sec:power-sample-size}. First we describe how we sampled the
survival processes under a specific hazard ratio. We then describe how we
estimated the maximum and expected sample size required to achieve a predefined
power (80\% in our case) for any of the test martingales that we considered
(that of the exact AV logrank, its Gaussian approximation, and the
prequential plugin variant). Finally, we explain how the same quantitiees for the
classical logrank test and the O'Brien-Fleming procedure were obtained.

%
In order to simulate the order in which the events in a survival processes
happens, we used the sequential-multinomial risk-set process from
Section~\ref{sec:safe-logrank-test-1}. As before, we consider the general
testing problem with \( \theta_{0} = 1 \) and a minimal clinically relevant
effect size \( \theta_{1} < 1 \), and we denote the true data generating
parameter by \( \theta \), typically, \( \theta \leq \theta_{1} \). Under
\( \theta \), the odds of the next event at the $i$\textsuperscript{th} event
time happening in Group $B$ are $\theta \bar{y}_{(i)}^B : \bar{y}^A_{(i)}$---the
odds change at each time step. %
%
%
Thus, simulating in which group the next event happens only takes a biased coin
flip. For the problem of testing \eqref{eq:two-sided_testing_problem} with
\( \theta_{0} \) we fix the tolerate a type-I error to $\alpha = 0.05$ and the
type-II error to \( \beta=0.2 \). For each test martingale $S_{\theta_0}^{q}$ of
interest we first consider the stopping rule
$\tau^{q} = \inf\{ k : S_{\theta_0,(k)}^{q} \geq 1/\alpha\}$, that is, we stop
as soon as $S_{\theta_0, (i)}^{q} $ crosses the threshold $1/\alpha$. Recall
that in the worst case, \( \theta = \theta_{1} \) the expected stopping time
\( \tau^{q} \) is lowest when we use \( S_{\theta_{0}, (k)}^{\theta_{1}} \), see
Appendix~{\ref{app:math-optimality}}.
%

To estimate the maximum number of events needed to achieve a predefined power
with a given test martingale, we turned our attention to a modified stopping
rule $\tilde{\tau}^{q}$. Under $\tilde{\tau}^{q}$ we stop at the first of two
moments: either when our test martingale $ S_{\theta_0, (k)}^{q}$ crosses the
threshold $1/\alpha$ (i.e., at $\tau$) or once we have witnessed a predefined
maximum number of events $n_{\max}$. More compactly, this means using the
stopping rule $\tilde{\tau}^{q}$ given by
$\tilde{\tau}^{q} = \min (\tau^{q}, n_{\max})$. In those cases in which the test
based on the stopping rule $\tau^{q}$ achieves a power higher than $1-\beta$, a
maximum number of events $n_{\max}$ smaller than the initial size of the
combined risk groups can be selected to achieve approximate power $1-\beta$
using the rule $\tilde{\tau}^{q}$.

A quick computation shows that $n_{\max}$ has the following property: it is the
smallest number of events $n$ such that stopping after $n$ events has
probability smaller than $1-\beta$ under the alternative hypothesis, that is,
%
%
\begin{equation*}
  \mathbf{P}_{\theta}\{\tau^{q} \geq n \} \leq 1-\beta.
\end{equation*}
More succinctly, $n_{\mathrm{max}}$ is the (approximate) $(1-\beta)$-quantile of the stopping time $\tau^{q}$, which can be estimated experimentally in a straightforward manner.

To estimate $n_{\max}$ for an initial risk set sizes $m_1,m_0$,
we sampled $10^4$ realizations
of the survival process (under $\theta$) using the method described at the
beginning of this section. This allowed us to obtain the same number of
realizations of the stopping time $\tau^{q}$. We then computed the
$(1-\beta)$-quantile of the simulated first passage time distribution of $\tau^{q}$, and
reported it as an estimate of the number of events $n_{\max}$ in the `maximum'
column in \autoref{fig:simulation_expected_vs_worst}.

We assessed the uncertainty in the estimation $n_{\mathrm{max}}$ using the
bootstrap. We performed 1000 bootstrap rounds on the sampled empirical
distribution of $\tau^{q}$, and found that the number of realizations that we
sampled ($10^4$) was high enough so that plotting the uncertainty estimates was
not meaningful relative to the scale of our plots. For this reason we omitted
the error bars in \autoref{fig:simulation_expected_vs_worst} and
\autoref{fig:simulation_sampleSize_misspecified}.

In the ``mean'' column of \autoref{fig:simulation_expected_vs_worst} and
\autoref{fig:simulation_sampleSize_misspecified} we plot an estimate of the
expected number of events $\tilde{\tau}^{q} = \min (\tau^{q}, n_{\max})$. For
this, we used the empirical mean of the stopping times that were smaller than
$n_{\max}$ on the sample that we obtained by simulation, with 20\% of the
stopping times being $n_{\max}$ itself. In the ``conditional mean'' column, we
plot an estimate of $\tilde{\tau}^{q} \mid \tilde{\tau}^{q}< n_{\max}$, i.e.,
the stopping time given that we stop early (and hence reject the null).

For comparison, we also show the number of events that one would need under the
Gaussian non-sequential approximation of \citet{schoenfeld_asymptotic_1981}, and
under the continuous monitoring version of the O'Brien-Fleming procedure. In
order to judge \citeauthor{schoenfeld_asymptotic_1981}'s approximation, we
report the number of events required to achieve 80\% power. This is equivalent
to treating the logrank statistic as if it were normally distributed, and
rejecting the null hypothesis using a $z$-test for a fixed number of events. The
power analysis of this procedure is classic, and the number of events required
is $n_{\max}^{S} = 4(z_\alpha + z_\beta)^2 / \log^2\theta_1$, where $z_\alpha$,
and $z_\beta$ are the $\alpha$, and $\beta$-quantiles of the standard normal
distribution. In the case of the continuous monitoring version of
O'Brien-Fleming's procedure, we estimated the number of events $n_{\max}^{OF}$
needed to achieve 80\% as follows. For each experimental setting
$(m^A,m^B, \theta)$, we generated $10^4$ realizations of the survival process
under \( \theta \) and computed the corresponding trajectories of the logrank
statistic. For each possible value $n$ of $n_{\max}^{OF}$, we computed the
fraction of trajectories for which the O'Brien-Fleming procedure correctly
stopped when used with the maximum number of events set to $n$. We report as an
estimate of the true $n_{\max}^{OF}$ the first value of $n$ for which this
fraction is higher than 80\%, our predefined power.

\section{Covariates: the full Cox Proportional Hazards $E$-Variable}
\label{app:extensions}
We extend the AV logrank test to the situation when time-dependent covariates
are present, as done in Section~\ref{sec:safe-logrank-test-1} with the same
notation used there.
Assume now the presence of $d$ covariates and let, for each participant $i$,
$\mathbf{z}_t^i = (z_{t,1}^i, \dots, z_{t,d}^i)$ be the covariate vector
consisting left-continuous time-dependent covariates
$z_{t,1}^{i}, \dots, z_{t,d}^{i}$. Denote by $\mathbf{z}_{(k)}^i$ the value of
the covariates of participant \( i \) at the time $T_{(k)}$ when the $k$th event
is witnessed. We let random variable $I_{(k)}$ denote the index of the patient
to which the $k$th event happens, and consider the extended process
$I_{(1)}, I_{(2)}, \dots$ where the information that is available at time $T_{(k)}$
is, $I_{(1)}, I_{(2)}, \dots, I_{(k)}$, and $z_{(1)}, \dots, z_{(k)}$.
%
%
The conditional partial likelihood underlying the process is now denoted
$\mathbf{P}_{\beta, \theta}$ with $\theta > 0$,
$\boldsymbol{\beta} \in \reals^d$, and $\beta_{\theta} = \ln \theta \in \reals$,
defined as follows:
\begin{align*}\label{eq:coxmodel}
  &\begin{multlined}
    \mathbf{P}_{\boldsymbol{\beta},\theta} \{
    I_{(k)} = i
    \mid
    \mathbf{z}_{(l)}^j , y_{(l)}^{j}; \ j = 1, \dots, n; \ l = 1, \dots, k
    \} :=\\
    P_{\beta,\theta}\{ I_{(k)} \mid \mathbf{z}_{(k)}^i, y_{(k)}^i; \ i = 1,
    \dots, n\}, \text{ and }
  \end{multlined}\\
  &
    \begin{multlined}
      \mathbf{P}_{\boldsymbol{\beta},\theta}\{
      I_{(k)} = i \mid \mathbf{z}_{(k)}^i , y_{(k)}^i; \ i = 1, \dots, n
      \}
      := \\
      p_{\boldsymbol{\beta},\theta, (k)}( \ i \ )
      :=
      \frac{
        y^i_{(k)}
        \exp\paren{\inner{\boldsymbol{\beta}}{\mathbf{z}^{i}_{(k)}} + g^i\beta_\theta}
      }{
        \sum_{j \in \mathcal{R}_{(k)}}
        \exp\paren{\inner{\boldsymbol{\beta}}{\mathbf{z}^{j}_{(k)}} + g^i\beta_\theta}
      },
    \end{multlined}
\end{align*}
This is consistent with Cox' (\citeyear{cox_regression_1972}) proportional
hazards regression model: the probability that the $i$th participant witnesses
an event, assuming he/she is still at risk, is proportional to the exponentiated
weighted covariates, with group membership being one of the covariates. In case
$\boldsymbol{\beta}= 0$, this is easily seen to coincide with the definition of
$\mathbf{P}_{\theta}$ via \eqref{eq:likelihood_per_time} with
$\theta = \rme^{\beta_\theta}$.

\subsection{$E$-Variables and Martingales}
Let $\mathbf{W}$ be a prior distribution on $\beta \in \reals^d$ for some
$d > 0$. ($\mathbf{W}$ may be degenerate, i.e., put mass one on a specific
parameter vector $\beta_1$). For each such $\mathbf{W}$, we let
$q_{\mathbf{W},\theta, (k)}$ be the probability distribution on
$\mathcal{R}_{(k)}$ defined by
\begin{align*}
  q_{\mathbf{W},\theta, (k)}( \ i  \ ) :=
  \int p_{\boldsymbol{\beta},\theta,(k)}( \ i \ ) \rmd \mathbf{W}(\boldsymbol{\beta}).
\end{align*}
Consider a measure $\rho$ on $\reals^d$ (e.g., Lebesgue or some counting measure)
and we let $\mathcal{W}$ be the set of all distributions on $\reals^d$ which
have a density relative to $\rho$, and $\mathcal{W}^{\circ} \subset \mathcal{W}$
be any convex subset of $\mathcal{W}$ (we may take
$\mathcal{W}^{\circ} = \mathcal{W}$, for example). We define
$\tilde{q}_{\leftarrow \mathbf{W},\theta_0}$ to be the {\em reverse information
  projection\/} \citep{li_estimation_1999} (RIPr) of
$q_{\mathbf{W},\theta, (k)}$ on
$\{q_{\mathbf{W},\theta_0,(k)}: \mathbf{W} \in \mathcal{W}^{\circ} \}$, defined
as the probability distribution on $\mathcal{R}_{(k)}$ such that
\begin{multline*}
  \KL( q_{\mathbf{W},\theta_1, (k)}  \|
  \tilde{q}_{\leftarrow \mathbf{W},\theta_0, (k)}  ) =\\
  \inf_{\mathbf{W}^{\circ} \in \mathcal{W}^{\circ}} \ \KL( q_{\mathbf{W},\theta_1, (k)}  \|
  q_{\mathbf{W}^{\circ},\theta_0, (k)} ).
\end{multline*}
We know from \citet{li_estimation_1999} and \citet{grunwald_safe_2020} that
$\tilde{q}_{\leftarrow \mathbf{W},\theta_0,(k)}$ exists for each $k$.
\citet{grunwald_safe_2020} show, in the context of $E$-variables for
$2 \times 2$ contingency tables, that the infimum in the previous display is in
fact achieved by some distribution $\mathbf{W}^{\star}$ with finite support on
$\reals^d$ if the random variables $y^1_{(k)}, \dots, y^m_{(k)} $ constituting
our random process have a finite range. For given hazard ratios
$\theta_0,\theta_1 > 0$, let
\begin{equation}
\label{eq:likelihoodRatiob}
  R_{\mathbf{W},\theta_0, (k)}^{\theta_1}
  =
  \frac{
    q_{\mathbf{W},\theta_1, (k)}( I_{(k)})
  }{
    q_{\leftarrow \mathbf{W},\theta_0, (k)}( I _{(k)})
  }
\end{equation}
be our analogue of \eqref{eq:one-outcome-evariable}.

%
%
\begin{theorem}[Corollary of Theorem 1 from \citet{grunwald_safe_2020}]
  \label{thm:infproj}
  For every prior $\mathbf{W}$ on $\reals^d$, for all
  $\boldsymbol{\beta }\in \reals^d$,
  \begin{multline*}
    \mathbf{E}_{\boldsymbol{\beta}, \theta_0}[
    R_{\mathbf{W},\theta_0, (k)}^{\theta_1}
    \mid
    \mathbf{z}_{(l)}^i , y_{(l)}^{i}; \ i = 1, \dots, m; \ l = 1, \dots, k
    ]
    = \\ \sum_{i \in \mathcal{R}_{(k)}}
    q_{\boldsymbol{\beta},\theta_0,(k)}(i )
    \frac{
      q_{\mathbf{W},\theta_1,(k)}( i )
    }{
      q_{\leftarrow \mathbf{W},\theta_0,(k)}(i)
    } \leq 1
  \end{multline*}
  so that $R_{\mathbf{W},\theta_0, (k)}^{\theta_1}$ is an $E$-variable
  conditionally on $\mathbf{z}_{(l)}^i , y_{(l)}^{i}$
  with$ \ i = 1, \dots, m; \ l = 1, \dots, k$.
\end{theorem}

Note that the result does not require the prior $\mathbf{W}$ to be well
specified in any way: under any $(\boldsymbol{\beta},\theta_0)$ in the null
distribution, even if $\boldsymbol{\beta}$ is completely disconnected to
$\mathbf{W}$, $R_{\mathbf{W},\theta_0, (k)}^{\theta_1}$ is an $E$-variable
conditional on past data.

In particular, since the result holds for arbitrary priors, it holds, at the
$k$th event time, for the Bayesian posterior
$ \mathbf{W}_{k+1} = \mathbf{W}_1 \mid \mathbf{z}_{(l)}^i , y_{(l)}^{i}; \ i =
1, \dots, m; \ l = 1, \dots, k$, based on arbitrary prior $\mathbf{W}_1$ with
density $w_1$, i.e., the density of $\mathbf{W}_{k+1}$ is given by
\begin{align}
  \nonumber
  w_{k+1}(\boldsymbol{\beta}) \propto \prod_{l\leq k}
  q_{\boldsymbol{\beta},\theta,(l)} (I_{(l)} ) w_1(\boldsymbol{\beta}).
\end{align}
%
%
In parallel to the discussion in Section~\ref{sec:optimality}, we can therefore,
for each prior $\mathbf{W}_1$, construct a test martingale
$S_{k} := \prod_{l\leq k} R_{\mathbf{W}_l,\theta_0,(l)}^{\theta_1}$ that
``learns'' $\boldsymbol{\beta}$ from the data, analogously
to~\eqref{eq:bayesian}, and computes a new RIPr at each event time $k$.

\subsection{Finding the RIPr}
While it is not clear how to calculate the RIPr
$q_{\leftarrow \mathbf{W},\theta_0,(k)}$ in general, it can be well approximated
with the efficient algorithm design by \citet{li_estimation_1999} and
\citet{li_mixture_1999}. Their algorithm is computationally feasible as long as
we restrict $\mathbf{W}^{\circ}_\delta$ to be the set of all priors $\mathbf{W}$
for which
$\min_{i \in \mathcal{R}_{(k)}} q_{\mathbf{W},\theta_0,(k)}( i ) \geq \delta$,
for some $\delta > 0$. In that case, when run for $M$ steps, the algorithm
achieves an approximation error of $O(\ln (1/\delta)/M)$, where each step is
linear in the dimension $d$. Since the approximation error is logarithmic in
$1/\delta$, we can take a very small value of $\delta$, which makes the
requirement less restrictive. Exploring whether the Li-Barron algorithm really
allows us to compute the RIPr for the Cox model, and hence
$R_{\mathbf{W}_k,\theta_0,(k)}^{\theta_1}$ in practice, is a major goal for
future work.
%
%

%
%
\subsection{Ties}
Without covariates, our $\E$-variables allow for ties correspond to a likelihood
ratio of Fisher's noncentral hypergeometric distributions (see
Section~\ref{sec:ties-1}), the situation is not so simple in the presence of
covariates. Although deriving the appropriate extension of the noncentral
hypergeometric partial likelihood is possible, one ends up with a
hard-to-calculate formula \citep{peto_discussion_1972}. Various approximations
have been proposed in the literature
\citep{cox_regression_1972,efron_efficiency_1977}. In case these preserve the
$E$-variable and martingale properties, they would retain type-I error
probabilities under optional stopping and we could use them without problems. We
do not know whether this is the case however; for the time being, we recommend
handling ties by putting the events in a worst-case order, leading to the
smallest values of the $\E$-variable of interest, as this is bound to preserve
the type-I error guarantees.

\section{Gaussian AV logrank test}
\label{app:gaussian_details}
In this section we derive the Gaussian AV logrank test of
Section~\ref{sec:appr-safel-test}, and investigate the validity of the Gaussian
approximation. In Appendix~\ref{app:safety-only-balanced}, we show that this
approximation is only valid when the allocation of participants to each group under
investigation is balanced, that is, when $m^A = m^B$. In
Appendix~\ref{app:sample-size} we investigate numerically the sample size needed
to reject the null hypothesis under both the exact AV logrank test and its
Gaussian approximation.

We start with the derivation of \eqref{eq:gaussian_safelogrank}. For this we
use (local) asymptotic normality of the $Z$-score \eqref{eq:logrank_z_score}. Under
the null distribution, $Z_k$ from \eqref{eq:logrank_z_score} has an asymptotic
standard Gaussian distribution. Under any alternative distribution under which
the hazard ratio is $\theta$, \citet{schoenfeld_asymptotic_1981} showed that, in
the absence of ties, the $Z$-statistic also follows a Gaussian distribution with
unit variance, but this time with mean $\mu_1^\star$ given by
\begin{equation*}
  \mu_1^\star
  =
  \frac{
    \sum_{i\leq k} E^B_i(1 - E^B_i)
  }{
    \sqrt{
      \sum_{i\leq k}
      E^B_i(1 - E^B_i)
    }
  }
  \log(\theta).
\end{equation*}
Note that $\mu_1^\star$ depends on more than the summary statistic $Z_k$.
In the case that the number of observed events is
much smaller than the initial risk set sizes, the mean $\mu_1^\star$ under the
alternative can be further approximated by
\begin{equation}
  \label{eq:compareschoenfeld}
  \mu_1^\star \approx \sqrt{\bar{N}_k} \mu_1 = \sqrt{\bar{N}_k} \sqrt{\frac{m^B m^A}{(m^B + m^A)^2}} \log(\theta),
\end{equation}
where $\bar{N}_k$ is the total number of observations up until time $t_k$, and
the resulting approximation only depends on summary statistics. It is exactly
this value $\mu_1$ that we use in the Gaussian AV logrank test. The asymptotic
result of \citeauthor{schoenfeld_asymptotic_1981} relies on two conditions: (1)
that the hazard ratio $\theta_1$ under the alternative is close enough to one so
that a first-order Taylor approximation around $\theta_0 = 1$ is adequate; (2)
that the expected number of events $E^B_{k}$ stays approximately constant over
time, that is, close to the initial allocation proportion
$E_1^B = m^B / (m^B + m^A)$. This indicates that the asymptotic approximation is
reasonable for values of $\theta_1$ close to $1$ and the initial risk sets are
both large in comparison to the number of events witnessed. Notice that in this
regime of large risk sets the multiplicity correction in $V_k$ is also
negligible.

\begin{figure*}[t!]
\centering
\includegraphics[scale = 1]{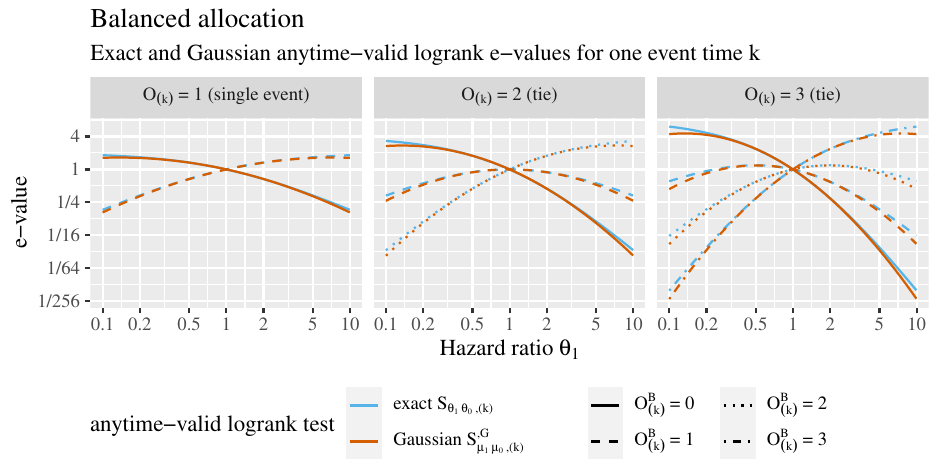}
\caption{For balanced allocation ($m^A = m^B$) $S'^{G}_1$ is very similar to
  $S_{(1)}$ when $0.5 \leq \theta_1 \leq 2$. Here $\theta_0 = 1$, $\mu_0 = 0$,
  and $\mu_1 = \mu_1(\theta_1)$ as in in \eqref{eq:compareschoenfeld}. Note that
  both axis are logarithmic. \label{fig:exactGaussianEvalues}}
\end{figure*}

This raises the question whether a sequential Gaussian approximation is sensible
for the logrank statistic--- a priori it is not at all clear whether Schoenfeld's
asymptotic fixed-sample result has a nonasymptotic counterpart. Define the the
logrank statistic per observation time
\begin{equation*}
  Z_i = \frac{O^B_i - E^B_i}{\sqrt{V^B_i}}.
\end{equation*}
We investigate whether the exact AV logrank statistic behaves similarly to the
Gaussian likelihood ratio
\begin{multline*}
  S_{k}^{'G}
  =
  \prod_{i\leq k} \frac{\phi_{\mu_1\sqrt{O_i}}(Z_i)}{\phi_{\mu_0}(Z_i)}
  =\\
  \exp\paren{
    -\frac12
    \sum_{i\leq k}
    \bracks{
      O_i\mu_1^2 - 2\mu_1\sqrt{O_iZ_i}
    }
  }
\end{multline*}
for $\theta_{0} = 1$ we have $\mu_0 = 0$,
$\mu_1 = \log(\theta)\sqrt{m^Bm^A / (m^A + m^B)^2}$, and $\phi_{\mu}$ is the
Gaussian density with unit variance and mean $\mu$. Note that the statistic
still depends on elements of the full data set; more approximations are needed.
Write the Gaussian densities, and use that in the limit of large risk sets
$p_i^B \approx m^B / (m^A + m^B)$ and that consequently
$V_i \approx \sqrt{O_i \frac{m^Am^B}{(m^A + m^B)^2}}$. This approximations valid
under Schoenfeld's second assumption. With these approximations at hand, the
$Z$-statistic is approximated by
\begin{equation*}
  Z_k
  \approx
  \frac{\sum_{i\leq k} \bracks{O^B_i - E^B_i}}{\sqrt{O_i \frac{m^Am^B}{(m^A + m^B)^2}}}
\end{equation*}
and consequently
\begin{equation*}
  {S'}^G_{k} \approx S^G_k
  \approx
  S_{k}^{G}
  =
  \exp\paren{
    -\frac12 \bar{N}_k\mu_1^2 + \sqrt{\bar{N}_k}\mu_1 Z_k},
\end{equation*}
where $S_k^G$ is as in \eqref{eq:gaussian_safelogrank}. 
In Figure~\ref{fig:exactGaussianEvalues} we
show, in case of balanced allocation, that the Gaussian approximation $S_k^G$ a
single event time from the Gaussian approximation are very similar to the exact
\( S_{\theta_{0},(k)}^{\theta_{1}} \) for alternative hazard ratios $\theta_1$ between 0.5 and 2.
%
%

\subsection{Safety only for balanced allocation}
\label{app:safety-only-balanced}
\begin{figure*}[t]
\centering
\includegraphics[scale = 1]{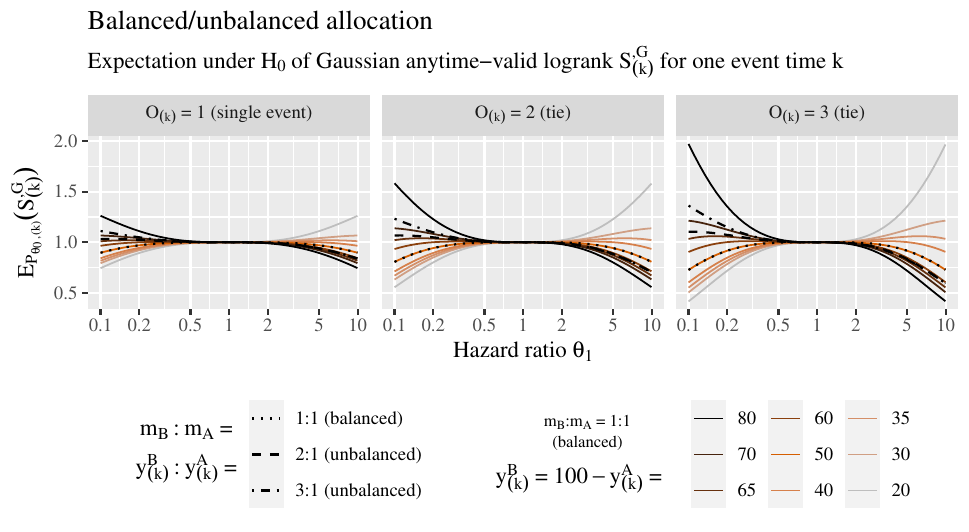}
\caption{Expected value  of the
  increments of the Gaussian AV logrank statistic as a function of the hazard
  ratio $\theta_1$. For balanced allocation $R^G_i$ is an $E$-variable, but it
  is not for unbalanced allocation. The risk set can also start out balanced but
  become unbalanced; this is unlikely under the null hypothesis (see
  Appendix~\ref{app:safety-only-balanced}). Note that the x-axis is
  logarithmic.\label{fig:expectationGaussian}} 
\end{figure*}

In order to assess whether the Gaussian AV logrank test is indeed AV, that
is, whether the type-I error guarantees holds, we inspect whether the
expected value of each of its multiplicative increments is bellow 1. In relation
to our discussion in Section~\ref{sec:optimality}, this would imply that all
multiplicative increments are conditional $E$-variables and that the resulting
test is, at least approximately, a test martingale.
\autoref{fig:expectationGaussian} shows the expectation of these increments as a
function of the hazard ratio for several initial allocation ratios. In case of
balanced 1:1 allocation $S^G_k$ is an $E$-variable, since its expectation is $1$
or smaller. However, in case of unbalanced 2:1 or 3:1 allocation and designs
with hazard ratio $\theta_1 < 1$, $S^G_k$ is not an $E$-variable. Of course,
even if the initial allocation is balanced, it can become unbalanced.
%
%
\autoref{fig:expectationGaussian} shows that in case of designs outside the
range $0.5 \leq \theta_1 \leq 2$ the deviations from expectation $1$ can be
problematic. Hence we do not recommend to use the Gaussian approximation on the
logrank statistic for unbalanced designs and designs for $\theta_1 < 0.5$ or
$\theta_1 > 2$. For balanced designs with $0.5 \leq \theta_1 \leq 2$, we found
that in practice they are safe to use, the reason being that scenarios in which
the allocation becomes highly unbalanced after some time (e.g.
$y_i^B = 80, y^A_i = 20$) are extremely unlikely to occur under the null.


\subsection{Sample size}
\label{app:sample-size}
\begin{figure*}[h!]
  \centering \includegraphics[scale = 1]{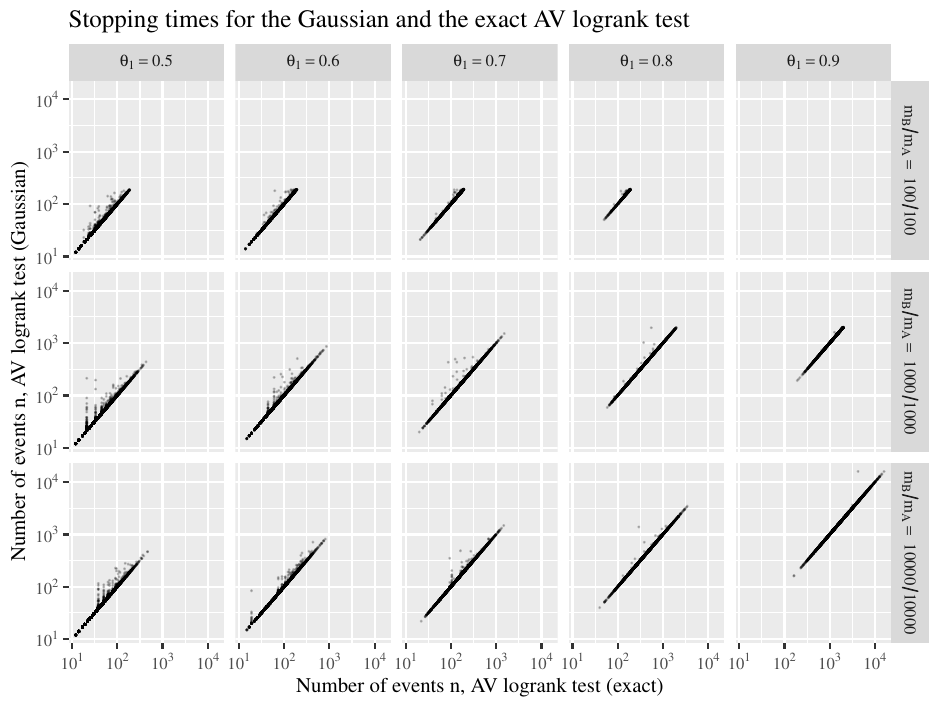}

  \caption{Stopping times for the Gaussian and exact AV logrank tests under
    continuous monitoring (no ties) with threshold $1 / \alpha = 20$. The
    stopping times under the Gaussian approximation often coincide with the
    exact ones, and are often more conservative (see
    Appendix~\ref{app:sample-size}). Note that both axes are
    logarithmic. \label{fig:exactGaussianStoppingTime}}.

\end{figure*}

%
%
In this section we compare the stopping time distribution \( \tau^{G} :=  \inf\{k : \xi^G_k = 1\} \) of the Gaussian approximation to that of \( \tau = \inf \{ k : \xi_k = 1\} \). %
%
We use tests with tolerable type I error $\alpha = 0.05$, thus, the threshold $ 1/\alpha = 20$ for both tests. In the previous section we showed that
the Gaussian approximation to the AV logrank statistic is valid when the initial
allocation is 1:1 and for values $0.5 \leq \theta_1 \leq 2$, where $\theta_1$ is
the hazard ratio under the alternative. In these scenarios, we simulate a
survival process from a distribution according to which the true data generating hazard ratio is $\theta = \theta_1$ and sampled realizations $\tau^G$ and $\tau$ for the same data set.
The results of the simulation are shown in
Figure~\ref{fig:exactGaussianStoppingTime}, where we plot the realizations of
$\tau^G$ against those of $\tau$. We see that in most cases both tests reject at
the same time $\tau^G = \tau$, and that the approximation becomes better as
$\theta_1$ moves closer to $\theta_0 = 1$ (Schoenfeld's assumption 1). When both
tests do not reject at the same time, the Gaussian approximation errs on the
conservative side. The deviations from the constant large and balanced risk set
do not seem to occur often for this range of hazard ratios. After all, the risk
set needs to be large to observe the number of events to detect hazard ratios in
the range $0.5 \leq \theta_1 \leq 2$.

\end{document}